\documentclass[aps,prb,reprint,superscriptaddress,amsmath,amssymb,floatfix,longbibliography]{revtex4-2}

\usepackage{graphicx}
\usepackage{bm}
\usepackage{booktabs}
\usepackage{dcolumn}
\usepackage[protrusion=true,expansion=true]{microtype}

\usepackage{xcolor}
\graphicspath{{figures/}}

\begin{document}

\title{Electron--phonon cooling noise in monolayer and bilayer graphene}
\author{Muhammad Waqas Haseeb}
\email{202090116@ueau.ac.ae}
\affiliation{Department of Physics, United Arab Emirates University, P.O. Box 15551 Al-Ain, United Arab Emirates}
\author{Edvin G. Idrisov}
\email{edvin.idrisov@ueau.ac.ae}
\affiliation{Department of Physics, United Arab Emirates University, P.O. Box 15551 Al-Ain, United Arab Emirates}


\begin{abstract}
We study the  electron--phonon heat-current fluctuations in monolayer and Bernal bilayer graphene coupled to longitudinal acoustic phonons. 
To do this, we construct a symmetric heat-current operator and evaluate the non-equilibrium current--current correlator at finite frequency for unequal electron and phonon temperatures. The heat-current vertices then involve the probe energy combined with twice the phonon energy, and the nonequilibrium spectral density must be calculated directly rather than inferred from a dynamic thermal fluctuation--dissipation relation. Further, we obtain full analytical expressions in the Bloch--Gr\"uneisen (BG) and Equipartition (EP) regimes, derive their low-frequency limits, and compare them with a complete doped-intraband on-shell calculations for monolayer graphene. The numerical results connect the two temperature regimes and show how finite-frequency noise changes out of equilibrium. For representative parameters, the corrections become percent-level in the upper-gigahertz to sub-terahertz range. It is worth mentioning that in equilibrium, the results satisfy the zero-frequency thermal fluctuation--dissipation relation(FDT) and reproduce the well established cooling-power and thermal-conductance laws. 
\end{abstract}


\maketitle

\section{Introduction}
\label{sec:introduction}

Electron--phonon energy relaxation determines the electronic temperature in
many mesoscopic devices based on two-dimensional materials \cite{Novoselov2004,Novoselov2005,KaasbjergCooling2014}. It controls how quickly hot electrons return to the
lattice temperature and therefore affects the physical mechanisms such as hot-electron transport, bolometric
response, calorimetric detection, and the recovery time of thermal detectors.
At low temperature, micron-scale graphene devices can have a very small
electronic heat capacity, so weak absorbed power can appreciably raise their electronic temperature \cite{Fong2012,Fong2013,Betz2012,Gasparinetti2015,Saira2016,Yan2012,Vora2012,Kokkoniemi2020}.

The coupling of graphene electrons to acoustic phonons has been studied extensively both in quasi-particle transport and cooling contexts
\cite{Hwang2008,Kubakaddi2009,Bistritzer2009,Tse2009,Viljas2010,Efetov2010,Borysenko2010,Kaasbjerg2012,Sohier2014}. To be specific, for long-wavelength longitudinal acoustic phonons, the dominant interaction is
commonly described by a deformation potential. This coupling results in two
important temperature regimes. At low temperature, only phonons with momenta
well below twice the Fermi wave vector are thermally active, and it is known as
Bloch--Gr\"uneisen (BG) regime \cite{Hwang2008,Viljas2010}. In the opposite limit of high temperatures, the active acoustic phonons
have large classical occupations, and the scattering becomes approximately
elastic. This is known as the equipartition (EP) regime
\cite{Hwang2008,Viljas2010,Min2011,Kaasbjerg2012}. The average electron--phonon cooling power has been studied for monolayer
graphene (MLG) and Bernal bilayer graphene (BLG) in both regimes \cite{Viljas2010,McCann2006,McCann2013,Borysenko2011}. In particular, it is worth pointing out that in Ref.~\cite{Viljas2010} authors considered a quasiequilibrium state in which the electrons and
phonons have separate temperatures. They obtained the limiting analytical
cooling laws and evaluated the numerical crossover between them
\cite{Viljas2010}. Related experimental and theoretical studies have examined
cooling in doped graphene, hot-electron transport, and disorder-assisted
energy relaxation
\cite{Kubakaddi2009,Bistritzer2009,Tse2009,Fong2012,Fong2013,Betz2012,Betz2013,Graham2013,Song2012,ChenClerk2012,Baker2012,Baker2013,McKitterick2016,Laitinen2014,Borzenets2013}.

The average cooling power, however, gives only the mean rate at which energy is
transferred. It is the first moment of the energy-transfer process. A complete
description should also determine the fluctuations around this mean value.
Microscopically, the energy exchange consists of separate phonon-emission and
phonon-absorption events. These events remain random even when the average
heat current is stationary. The corresponding heat-current noise is the
second moment of the cooling process and is the focus of our work. Heat-current fluctuations have previously been considered in metallic
electron--phonon systems and in mesoscopic thermal links
\cite{Giazotto2006,Averin2010,Pekola2018,Karimi2021}. These studies show that finite-frequency heat
noise contains information that is absent from the average thermal
conductance. In particular, a simple frequency-dependent extension of the
classical thermal fluctuation--dissipation theorem (FDT) does not generally contain
all quantum contributions to heat-current noise
\cite{Averin2010,Pekola2018,Karimi2021}. As a next step, these studies motivate a direct microscopic calculation of the finite-frequency spectrum. As in previous works, we consider the clean MLG and BLG coupled to longitudinal acoustic phonons through an unscreened deformation potential. The
electron subsystem is assumed to remain internally thermal at an electron
temperature \(T_{\rm e}\), while the acoustic phonons remain internally thermal at a
phonon temperature \(T_{\rm ph}\). This two-temperature description is the
quasiequilibrium approximation used in the standard cooling-power theory
\cite{Viljas2010,Giazotto2006,Pekola2018}. The two temperatures are distinct, corresponding to a non-equilibrium state. This non-equilibrium condition results in energy exchange between electron and phonon subsystems. Namely, phonon emission transfers energy from electrons to phonons, whereas phonon absorption transfers energy in the opposite direction.
The average heat current measures the imbalance between the two processes, while the heat-current noise measures their combined fluctuations  (see Fig.~\ref{fig:schematic}). 

The remainder of the paper is organized as follows. In Sec.~\ref{sec:formulation of problem} we formulate the problem and discuss the main features of it. Sec.~\ref{sec:model} introduces the microscopic model and heat-current
operator. Sec.~\ref{sec:kernel} derives the leading in electron-phonon coupling current--current
correlator and identifies the physical exchange contribution.
Sec.~\ref{sec:zero_results} presents the zero-frequency MLG and BLG
results in the BG and EP regimes.
Sec.~\ref{sec:finite_results} develops the finite-frequency spectra for
equal and unequal electron and phonon temperatures.
Sec.~\ref{sec:numerics} presents the numerical crossover calculations,
finite-frequency comparisons, and experimental noise scales.
Sec.~\ref{sec:discussion} discusses the physical interpretation,
applicability, and limitations of the results. Detailed derivations and
numerical checks are collected in the appendices.

\section{Formulation of problem}
\label{sec:formulation of problem}

To study the fluctuations, we define the symmetrized heat-current noise spectrum, which is sum of two unsymmetrized terms
\begin{equation}
 S(\omega)
 =
 \frac{1}{2}
 \left[
 \mathcal{Z}(\omega)
 +
 \mathcal{Z}(-\omega)
 \right].
 \label{eq:intro_sym_noise}
\end{equation}
In case of time invariance under the consideration, each unsymmetrized noise is given by the Fourier transformation of two-point correlation function of heat currents
\begin{equation}
 \mathcal{Z}(\omega)
 =
 \int dt\,
 e^{i\omega t}
 \left\langle
 \delta \hat{J}(t)\delta \hat{J}(0)
 \right\rangle,
 \label{eq:intro_unsym_noise}
\end{equation}
where \(\delta \hat{J}=\hat{J}-\langle \hat{J}\rangle\) and \(\hat{J}(t)\) is the heat-current operator in the Heisenberg picture, which is defined later in Eqs.~(\ref{eq:current_definition}) and (\ref{eq:heat_current_operator}). We choose
the sign of \(\hat{J}\) so that a positive current corresponds to energy flowing
from the electrons to the phonons due to the fact that \(T_{\rm e} \geq T_{\rm ph} \). The quantity \(\langle \hat{J}\rangle\) is the
average heat current, while \(\delta \hat{J}(t)\) can be considered as the instantaneous fluctuation
around that average. The average is taken with respect to non-equilibrium density matrix which at $t \to -\infty$ is given by the product of equilibrium electron and phonon sub-systems \(\hat{\rho}_{\rm e} \otimes \hat{\rho}_{\rm ph}\) with corresponding temperatures \(T_{\rm e}\) and \(T_{\rm ph}\). The frequency \(\omega\) sets the time scale probed by the measurement. Positive and negative frequencies are physically different in the unsymmetrized spectrum because they describe the ability of the system to supply energy to a detector or absorb energy from it. In contrast, the Eq.~(\ref{eq:intro_sym_noise}) treats positive and negative frequencies equally and gives a spectrum that is even in frequency. At zero frequency, the symmetrized and unsymmetrized spectra coincide. Therefore, our work investigates how the spectrum given in Eq.~(\ref{eq:intro_sym_noise}) changes when the detector resolves fluctuations at finite frequencies, whether the zero-frequency noise recovers the known equilibrium thermal conductance, and how these results evolve across BG and EP regimes, both at equilibrium and non-equilibrium situations, when the electron and phonon temperatures differ.

\begin{figure}[t]
 \centering
 \includegraphics[width=\columnwidth]{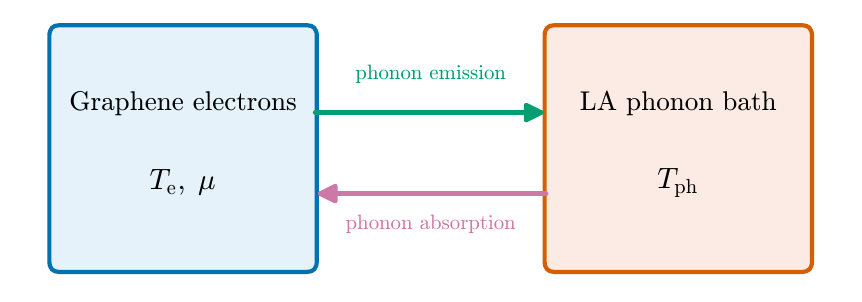}
 \caption{Schematic electron--phonon energy exchange in the quasiequilibrium two-temperature description. Phonon emission transfers
 energy from electrons to phonons, whereas phonon absorption transfers energy
 in the opposite direction. The average heat current measures the imbalance
 between the two processes, while the heat-current noise measures their
 combined fluctuations.}
 \label{fig:schematic}
\end{figure}

The physical distinction between the mean heat current and its noise can be
understood from phonon emission and absorption processes. In an emission event, an electron loses energy and creates a phonon. In an absorption event, an
electron gains energy by removing a phonon from the phonon bath. The average
heat current assigns opposite signs to these two events because they transfer
energy in opposite directions. The noise assigns the same sign to both events
because each event contributes positively to the variance of the transferred
energy. The average current therefore measures the imbalance between the emission and absorption rates. The zero-frequency noise measures their combined
fluctuations and weights each event by the square of the transferred phonon
energy. At equilibrium, the emission and absorption rates are equal. Their
difference vanishes, so the average current is zero, but their sum remains
finite, resulting in non vanishing heat-current noise. This physical picture is summarized in Fig.~\ref{fig:schematic}. The electron and phonon subsystems continue to exchange energy even when their temperatures are equal. Equilibrium means that the two opposite energy currents cancel on average, not that the
microscopic energy-transfer events stop. Moreover, even at zero temperature due to non-zero electron-phonon coupling the quantum fluctuations of heat current survive at finite frequency. 

Our calculations starts from the electron, acoustic-phonon, and
electron--phonon Hamiltonians. We construct a symmetric heat-current operator
from the rates of change of the electron and phonon subsystem energies. This
definition is closely related to the symmetric heat-current operator used in
finite-frequency electron--phonon noise calculations in normal metals and low-dimensional systems~\cite{Pekola2018,Averin2010,IdrisovThermalDrag2022}. We then evaluate the leading current--current correlation function in electron-phonon coupling by performing the complete Wick expansion. The fermionic part produces a direct
contraction and an exchange contraction. The direct contraction forces the
initial and final electron states to be identical. Momentum conservation then
requires the acoustic phonon momentum to be zero. A zero-momentum acoustic
phonon carries no energy, so this contribution vanishes. The exchange contraction describes a genuine electron--phonon scattering
event and gives the physical noise spectrum. It contains both phonon emission
and phonon absorption. Keeping both processes is essential at zero frequency.
Removing either one would give an incomplete rate sum and would prevent the
equilibrium noise from reproducing the thermal conductance correctly.

At equilibrium, the zero-frequency noise is related to the linear
electron--phonon thermal conductance. More precisely the noise equals twice the
Boltzmann constant multiplied by the square of the base temperature and by
the conductance. The thermal conductance, \(G_{\rm ep}(T)\), is defined as the response of the average
heat current to a small temperature difference between electron and phonon subsytems, \(\Delta T\), around base temperature, \(T\), and the corresponding equilibrium zero-frequency noise is
\(
S(0) = 2k_{\mathrm{B}}T^2G_{\mathrm{ep}}(T)\). We use this relation only at zero frequency, where it
provides a stringent check on the microscopic calculations \cite{Callen1951,Kubo1957,Averin2010,Pekola2018}. The zero-frequency calculation reproduces the established cooling-power laws
of monolayer and bilayer graphene \cite{Hwang2008,Kubakaddi2009,Viljas2010,McKitterick2016}. In the clean
BG regime, the MLG conductance is proportional to the absolute
chemical potential and to the third power of temperature (\( \propto T^3\)). Its equilibrium
noise is therefore proportional to the absolute chemical potential and to the
fifth power of temperature (\(\propto T^5\)). The BLG conductance has the same third-power temperature dependence in the BG regime, and its equilibrium noise again has a fifth-power temperature dependence. Its doping dependence is different, however. The BLG
coefficient is proportional to the inverse square root of the absolute
chemical potential. In contrast to MLG, this difference follows from the parabolic low-energy
dispersion of BLG and from the chemical-potential
dependence of its Fermi velocity \cite{Viljas2010,McCann2006,McCann2013,Borysenko2011,Min2011}. In EP regime, the active acoustic phonons are classically
occupied. For MLG, the thermal conductance is controlled by a fourth-order
electronic energy moment. This produces separate contributions from the
chemical potential and thermal broadening. For BLG, the parabolic electronic
dispersion leads to a temperature multiplied by a logarithmic phase-space
factor. As a result, MLG and BLG show different high-temperature and
near-neutrality behavior \cite{Hwang2008,Viljas2010,Min2011,Borysenko2011}. In equilibrium but at finite frequency, the detector introduces an additional energy scale, which survives even in case of zero temperature of whole system, which is the result of quantum fluctuations due to finite electron-phonon coupling. The
heat-current vertices then contain the square of the detector energy minus
twice the phonon energy and the square of the detector energy plus twice the
phonon energy. These factors show that finite-frequency noise probes the time
dependence of electron--phonon energy exchange rather than only its long-time
variance. Previous works have shown that finite-frequency heat-current noise can contain quantum contributions that are not determined by the ordinary dynamic thermal conductance \cite{Averin2010,Pekola2018,Karimi2021}. Direct evaluation is also necessary away from equilibrium because no single temperature characterizes both subsystems.

Away from equilibrium, at \(T_{\rm e} \geq T_{\rm ph}\), the electron temperature controls the available electron--hole phase space, while the phonon temperature controls the acoustic-phonon occupation. The resulting spectrum is therefore an independent non-equilibrium rate-sum observable. It should not be interpreted as a thermal conductance. Therefore, we derictly derive the finite-frequency spectra in BG and EP regimes for both MLG and BLG. In each regime, we compare the exact finite-frequency integral with its low-frequency approximation. This comparison identifies the frequency range in which the leading quadratic correction is reliable, following the general strategy used in finite-frequency electron--phonon noise studies \cite{Pekola2018,Karimi2021}. For doped MLG, we additionally evaluate a full intraband on-shell integral
without making either BG or EP reduction. This
calculation retains the exact scattering geometry, graphene chirality factor,
Fermi occupation, Bose occupation, and finite-frequency current vertices
within the clean acoustic deformation-potential model. It approaches the
BG result at low temperature and EP equipartition result at
high temperature, while describing the smooth crossover between them \cite{Hwang2008,Viljas2010,Kaasbjerg2012,Sohier2014}.

The numerical analysis includes the temperature dependence of the thermal
conductance and zero-frequency noise, the full
BG-to-EP crossover, normalized finite-frequency
spectra, exact-versus-asymptotic comparisons, nonequilibrium frequency maps,
and absolute device-scale noise estimates. For representative graphene
parameters, the finite-frequency correction is generally small at ordinary
low-gigahertz frequencies but can become measurable in the upper-gigahertz
and sub-terahertz ranges. Finite-frequency heat-current fluctuations are
particularly relevant to fast thermometry and calorimetric detection
\cite{Pekola2018,Fong2012,Fong2013,Gasparinetti2015,Saira2016,Yan2012,Vora2012,Kokkoniemi2020}. It is worth mentioning, that the present model is restricted to clean graphene coupled to
longitudinal acoustic phonons through an unscreened deformation potential.
The unscreened model is commonly used to describe intrinsic acoustic-phonon
scattering in graphene \cite{Hwang2008,Kubakaddi2009,Viljas2010,Efetov2010,Borysenko2010,Kaasbjerg2012}. Screening,
disorder-assisted supercollisions, flexural phonons, optical phonons,
substrate modes, and interface thermal resistance are outside the present
scope. Some of these mechanisms can dominate in different temperature,
density, disorder, or substrate regimes \cite{Viljas2010,Song2012,ChenClerk2012,Betz2013,Graham2013,Laitinen2014,Sohier2014}. Keeping
the model restricted allows the microscopic structure of the heat-current
noise and the relation between its limiting regimes to be identified clearly analytically.

\section{Theoretical model and heat-current operator}
\label{sec:model}

\subsection{Total Hamiltonian}

We consider electrons in graphene coupled to in-plane longitudinal acoustic phonons through the deformation potential.  The total Hamiltonian consists of three terms
\begin{equation}
 \hat{H}=\hat{H}_{\rm e}+\hat{H}_{\rm ph}+\hat{H}_{\rm e-ph}.
 \label{eq:H_total}
\end{equation}
The first term corresponds to free MLG or BLG Hamiltonian and is given by 
\begin{equation}
 \hat{H}_{\rm e}=\sum_{{\bf k},\alpha}\epsilon_{{\bf k}\alpha}
 \hat{c}_{{\bf k}\alpha}^{\dagger}\hat{c}_{{\bf k}\alpha},
 \label{eq:H_e}
 \end{equation}
where the electron with momentum \({\bf k}\) and band index \(\alpha\) has energy \(\epsilon_{{\bf k}\alpha}\). The operators \(c^\dagger\) and \(c\) create and annihilate electrons, respectively. The band index is \(\alpha=+1\) for the conduction band and \(\alpha=-1\) for the valence band. For MLG the spectrum is linear in momentum,
\(
\epsilon_{{\bf k}\alpha}=\alpha\hbar v_{\rm F} k
\), and \(v_{\rm F}\) is a Fermi velocity.
That is, conduction-band electrons have energy \(+\hbar v_{\rm F} k\), while valence-band states have energy \(-\hbar v_{\rm F} k\).
For BLG, the low-energy dispersion is parabolic:
\(
\epsilon_{{\bf k}\alpha}
=
\alpha \hbar^2k^2 /2m \) with effective mass
\(
m=\gamma_1 /2v_0^2
\), and \(v_0 \simeq v_{\rm F}\) is the intralayer band velocity of graphene and \(\gamma_1\) is the dominant interlayer hopping energy in BLG. It describes electron hopping between the carbon atoms that lie directly above and below one another. This bilayer approximation further is valid when the chemical potential, electron and phonon temperatures, and relevant frequency scale are all much smaller than the interlayer hopping, i.e.
\(
\max\left(
|\mu|,
k_{\rm B}T_{\rm e},
k_{\rm B}T_{\rm ph},
\hbar|\omega|
\right)
\ll \gamma_1.
\)
The second term is the phonon Hamiltonian
 \begin{equation}
 \hat{H}_{\rm ph}=\sum_{\bf q}E_q \hat{b}_{\bf q}^{\dagger}\hat{b}_{\bf q},
 \label{eq:H_ph}
 \end{equation}
 where \(E_q=\hbar v_{\rm s} q\) is the acoustic phonon energy, \(v_{\rm s}\) is the sound velocity of the longitudinal acoustic phonons. The third term is the electron-phonon interaction
 \begin{equation}
 \hat{H}_{{\rm e}\text{-}{\rm ph}}=\sum_{{\bf k}\alpha,{\bf p}\beta,{\bf q}}
 M_{{\bf p}{\bf k},{\bf q}}^{\beta\alpha}
 \hat{c}_{{\bf p}\beta}^{\dagger}\hat{c}_{{\bf k}\alpha}
 \left(\hat{b}_{\bf q}+\hat{b}_{-{\bf q}}^{\dagger}\right).
 \label{eq:H_ep}
\end{equation}
In the above Eq.~(\ref{eq:H_ep}) the product of electron annihilation and creation operators, \(\hat{c}_{\mathbf{p}\beta}^{\dagger}\hat{c}_{\mathbf{k}\alpha}\), removes an electron from the initial state \((\mathbf{k},\alpha)\) and places it in the final state \((\mathbf{p},\beta)\). The coefficient \(M_{\mathbf{p}\mathbf{k},\mathbf q}^{\beta\alpha}\) is the electron–phonon coupling matrix element, which contains the deformation-potential strength and the chirality overlap between the two electronic states.
The bosonic annihilation operator \(\hat{b}_{\mathbf q}\) corresponds to phonon absorption, i.e. an existing phonon is annihilated while the electron changes momentum. In contrast, the bosonic creation operator \(\hat{b}_{-\mathbf q}^{\dagger}\) corresponds to phonon emission, namely the electron loses energy and creates a phonon with momentum \(-\mathbf q\). The translational invariance imposes ${\bf p}={\bf k}+{\bf q}$.  The deformation-potential matrix element is known~\cite{Viljas2010} and may be written as
\begin{equation}
 \left|M_{{\bf p}{\bf k},{\bf q}}^{\beta\alpha}\right|^2
 =\frac{D^2\hbar q}{2A\rho_{\rm m} v_{\rm s}}F_{\alpha\beta}(\phi),
 \label{eq:matrix_element}
\end{equation}
where $A$ is the graphene area and $\rho_{\rm m}$ is the areal mass density of the corresponding mono- or bilayer system.  The Dirac chirality factors are~\cite{Viljas2010,Hwang2008}
\begin{align}
 F_{\alpha\beta}^{\rm MLG}(\phi)&=\frac{1+\alpha\beta\cos\phi}{2},
 \label{eq:chirality_mlg}\\
 F_{\alpha\beta}^{\rm BLG}(\phi)&=\frac{1+\alpha\beta\cos2\phi}{2},
 \label{eq:chirality_blg}
\end{align}
with $\phi$ the angle between the incoming and outgoing electron momenta. Furthermore, the two subsystems are assumed to be internally thermal, thus the distribution functions are given by Fermi-Dirac $f_{\rm e}(\epsilon;T_{\rm e})=1/[\exp[(\epsilon-\mu)/(k_{\rm B}T_{\rm e})]+1]$ for electrons and Bose-Einstein $n_{\rm ph}(E;T_{\rm ph})=1/[\exp[E/(k_{\rm B}T_{\rm ph})]-1]$ for phonons. This is the same quasiequilibrium framework used in the cooling-power theory of Ref.~\cite{Viljas2010}.  It is appropriate when electron--electron scattering maintains a Fermi distribution and phonon relaxation maintains a Bose distribution even though $T_e$ and $T_{\rm ph}$ differ.

\subsection{Heat-current operator}
\label{subsec:heat_current_operator}
We define the heat current to be positive when energy is transferred from the
electron subsystem to the phonon subsystem, thus \(T_{\rm e} \geq T_{\rm ph}\). The electron and phonon
Hamiltonians are denoted by \(\hat{H}_{\rm e}\) and \(\hat{H}_{\rm ph}\), respectively.  Their
Heisenberg time derivatives describe the instantaneous rates at which the
energies of the two subsystems change.  We use the symmetric heat current
definition
\begin{equation}
 \hat{J}=\frac{1}{2} \partial_t\left(\hat{H}_{\rm ph}- \hat{H}_{\rm e}\right),
 \label{eq:current_definition}
\end{equation}
where \(\partial_t \hat{H}_i=i[\hat{H},\hat{H}_i]/\hbar\) and \(\hat{H}\) is the total Hamiltonian, including the electron, phonon, and
electron--phonon interaction terms from Eq.~(\ref{eq:H_total}). The sign convention in Eq.~\eqref{eq:current_definition} has a direct physical
meaning, namely when electrons lose energy, \( \partial_t \langle \hat{H}_{\rm e} \rangle<0\), and when phonons gain the
same energy, \( \partial_t\langle  \hat{H}_{\rm ph} \rangle >0\).  Both contributions then make the average of \(\hat{J}\)
positive.  Reversing the convention would replace \(\hat{J}\) by \(-\hat{J}\).  The
average current would change sign, but the current noise, which contains a
product of two current operators, would remain unchanged, which one can see in next sections. The factor \(1/2\) treats the changes of the electron and phonon energies
symmetrically.  This is important because, at a finite observation frequency, the
interaction energy cannot in general be assigned completely to either
subsystem.  In the zero-frequency steady-state limit, the electron energy lost
and the phonon energy gained become equal.  Throughout this work, the
symmetric convention in Eq.~\eqref{eq:current_definition} is used
consistently.

Only the electron--phonon interaction contributes to the commutators in
Eq.~\eqref{eq:current_definition}.  The electron Hamiltonian commutes with the
free phonon Hamiltonian, and each free subsystem Hamiltonian commutes with
itself.  Evaluating the commutators gives
\begin{align}
 \partial_t \hat{H}_{\rm e}
 &=
 \frac{i}{\hbar}
 \sum_{{\bf k}\alpha,{\bf p}\beta,{\bf q}}
 M_{{\bf p}{\bf k},{\bf q}}^{\beta\alpha}
 \Delta_{kp}^{\alpha\beta}
 \hat{c}_{{\bf p}\beta}^{\dagger}\hat{c}_{{\bf k}\alpha}
 \left(
 \hat{b}_{\bf q}+\hat{b}_{-{\bf q}}^{\dagger}
 \right),
 \label{eq:Hedot}\\
 \partial_t  \hat{H}_{\rm ph}
 &=
 \frac{i}{\hbar}
 \sum_{{\bf k}\alpha,{\bf p}\beta,{\bf q}}
 M_{{\bf p}{\bf k},{\bf q}}^{\beta\alpha}
 E_q\,
 \hat{c}_{{\bf p}\beta}^{\dagger}\hat{c}_{{\bf k}\alpha}
 \left(
 \hat{b}_{\bf q}-\hat{b}_{-{\bf q}}^{\dagger}
 \right),
 \label{eq:Hphdot}
\end{align}
where to describe an electron transition, we introduce the energy difference
\begin{equation}
 \Delta_{kp}^{\alpha\beta}
 =
 \epsilon_{{\bf k}\alpha}
 -
 \epsilon_{{\bf p}\beta}.
 \label{eq:delta_energy}
\end{equation}
The vector \({\bf k}\) and band index \(\alpha\) label the initial electron
state, whereas \({\bf p}\) and \(\beta\) label the final electron state.
The corresponding band energies are
\(\epsilon_{{\bf k}\alpha}\) and
\(\epsilon_{{\bf p}\beta}\).  Thus,
\(\Delta_{kp}^{\alpha\beta}>0\) means that the electron loses energy during
the transition, while
\(\Delta_{kp}^{\alpha\beta}<0\) means that the electron gains energy. The phonon energy \(E_q\) is given in Eq.~(\ref{eq:H_ph}) and electron-phonon matrix element \(M_{{\bf p}{\bf k},{\bf q}}^{\beta\alpha}\) gives the amplitude for the
electron--phonon scattering process, Eq.~(\ref{eq:H_ep}). The different signs of the creation and annihilation operators in
Eq.~\eqref{eq:Hphdot} follow from their commutators with the phonon
Hamiltonian.  An annihilation operator removes the energy \(E_q\) from the
phonon subsystem, whereas a creation operator adds the energy \(E_q\).
By contrast, both terms enter Eq.~\eqref{eq:Hedot} with the same sign inside
the parentheses because both are coupled to the same electronic transition
operator; their physical distinction is imposed by the corresponding energy
shells.

Substituting Eqs.~\eqref{eq:Hedot} and \eqref{eq:Hphdot} into
Eq.~\eqref{eq:current_definition} gives the symmetric heat-current operator
\begin{align}
 \hat{J}
 &=
 \frac{i}{2\hbar}
 \sum_{{\bf k}\alpha,{\bf p}\beta,{\bf q}}
 M_{{\bf p}{\bf k},{\bf q}}^{\beta\alpha}
 \hat{c}_{{\bf p}\beta}^{\dagger}\hat{c}_{{\bf k}\alpha}
 \nonumber\\
 &\quad\times
 \left[
 \left(E_q-\Delta_{kp}^{\alpha\beta}\right)\hat{b}_{\bf q}
 -
 \left(E_q+\Delta_{kp}^{\alpha\beta}\right)
 \hat{b}_{-{\bf q}}^{\dagger}
 \right].
 \label{eq:heat_current_operator}
\end{align}
The full momentum and band sums, together with the Hermiticity properties of
the electron--phonon matrix element, ensure that
Eq.~\eqref{eq:heat_current_operator} represents a Hermitian physical observable. The first term in Eq.~\eqref{eq:heat_current_operator} contains
\(b_{\bf q}\) and therefore describes a process in which a phonon is
annihilated.  The electron absorbs the phonon energy and moves to a state of
higher energy.  On the corresponding zero-frequency energy shell, the
electron energy difference is negative and has magnitude \(E_q\).  The
coefficient \(E_q-\Delta_{kp}^{\alpha\beta}\) then has magnitude \(2E_q\). The second term contains \(b_{-{\bf q}}^{\dagger}\) and describes phonon
creation.  In this process the electron moves to a lower-energy state and
transfers energy to the phonon subsystem.  On the corresponding
zero-frequency energy shell, the electron energy difference is positive and
equal in magnitude to \(E_q\).  The coefficient
\(E_q+\Delta_{kp}^{\alpha\beta}\) again has magnitude \(2E_q\).

This observation is important for the noise calculation further.  The two current
vertices are squared in the heat current--current correlator.  At zero frequency,
both the emission and absorption contributions therefore acquire the same
energy weight, namely four times the square of the phonon energy.  The two
processes consequently enter the zero-frequency noise with equal energy
weights and must both be retained. At finite frequency, the energy provided to or absorbed from a detector shifts
the two energy-conservation conditions.  The two current vertices then become
the square of the detector energy minus twice the phonon energy and the square
of the detector energy plus twice the phonon energy.  The distinction between
the two coefficients in Eq.~\eqref{eq:heat_current_operator} is therefore the
microscopic origin of the nontrivial finite-frequency spectrum.

\section{Heat-current noise}
\label{sec:kernel}
The unsymmetrized spectrum is defined in Eq.~\eqref{eq:intro_unsym_noise}. In the leading order of electron-phonon coupling, the fermionic four-point function contains direct and exchange Wick contractions.  The direct contraction sets ${\bf p}={\bf k}$.  Momentum conservation then forces ${\bf q}=0$, and the acoustic phonon energy and both current vertices vanish, thus the direct contribution is identically zero.  The exchange contraction sets the two electron lines into a real scattering process and gives the occupation factor $f_{{\bf p}\beta}(1-f_{{\bf k}\alpha})$. After some algebra (see details in Appendix~\ref{app:wick}) we obtain the following expression for noise
\begin{widetext}
\begin{align}
 \mathcal{Z}(\omega)
 &=\frac{\pi g_d}{2\hbar}
 \sum_{{\bf k}\alpha,{\bf p}\beta,{\bf q}}
 \left|M_{{\bf p}{\bf k},{\bf q}}^{\beta\alpha}\right|^2
 f_{{\bf p}\beta}(T_{\rm e})\left[1-f_{{\bf k}\alpha}(T_{\rm e})\right]
 \nonumber\\
 &\quad\times\Big\{
 \left(\Delta_{kp}^{\alpha\beta}-E_q\right)^2
 \left[1+n_{\rm ph}(E_q,T_{\rm ph})\right]
 \delta\!\left(\hbar\omega-\Delta_{kp}^{\alpha\beta}-E_q\right)
 \nonumber\\
 &\qquad\quad+
 \left(\Delta_{kp}^{\alpha\beta}+E_q\right)^2
 n_{\rm ph}(E_q,T_{\rm ph})
 \delta\!\left(\hbar\omega-\Delta_{kp}^{\alpha\beta}+E_q\right)
 \Big\}.
 \label{eq:general_unsym_kernel}
\end{align}
\end{widetext}
Eq.\eqref{eq:general_unsym_kernel} is the leading contribution,
of second order in the electron--phonon coupling, to the unsymmetrized
heat-current noise. The electronic occupation factor
\(f_{{\bf p}\beta}(T_{ \rm e})[1-f_{{\bf k}\alpha}(T_{\rm e})]\) requires the initial
electron state \(({\bf p},\beta)\) to be occupied and the final state
\(({\bf k},\alpha)\) to be empty. The first factor is therefore the
probability of finding the initial electron, while the second is the Pauli
blocking factor for the final state. Both are evaluated at the electron
temperature \(T_{ \rm e}\).

The first term in Eq.~\eqref{eq:general_unsym_kernel} is associated with
phonon emission. Its statistical factor
\(1+n_{\rm ph}(E_q,T_{\rm ph})\) contains both spontaneous emission and
Bose-enhanced stimulated emission. The second term is associated with phonon
absorption and is proportional to the phonon occupation
\(n_{\rm ph}(E_q,T_{\rm ph})\), because absorption is possible only when a
phonon is already present in the mode. The phonon distribution is evaluated
at the phonon temperature \(T_{\rm ph}\), which may differ from \(T_{\rm e}\). The two Dirac delta functions impose energy conservation for the two
processes, including the energy \(\hbar\omega\) exchanged with a
finite-frequency probe. Consequently, positive and negative frequencies are
not generally equivalent in the unsymmetrized spectrum. The factors
\((\Delta_{kp}^{\alpha\beta}-E_q)^2\) and
\((\Delta_{kp}^{\alpha\beta}+E_q)^2\) are the squared heat-current vertices
generated by the symmetric current operator. They determine how strongly
each allowed scattering event contributes to the noise. At zero frequency, the emission shell fixes
\(\Delta_{kp}^{\alpha\beta}=-E_q\), whereas the absorption shell fixes
\(\Delta_{kp}^{\alpha\beta}=+E_q\). The two current vertices then acquire the
same value, \(4E_q^2\). Thus emission and absorption carry equal
energy-transfer weights in the zero-frequency noise and must both be retained.
At finite frequency, the two weights are different and become functions of
both the probe energy and the phonon energy. This difference is the
microscopic origin of the nontrivial frequency dependence of the
heat-current spectrum. The spectrum in Eq.~\eqref{eq:general_unsym_kernel} is not generally even in
\(\omega\). The symmetrized spectrum used below is obtained by averaging the
positive- and negative-frequency contributions according to
Eq.~\eqref{eq:intro_sym_noise}.

At zero frequency the delta functions in Eq.~(\ref{eq:general_unsym_kernel}) impose that $\Delta= \pm E_q$. In both cases the squared current vertex becomes $4E_q^2$.  In intraband notations one gets
\begin{align}
 S(0)&=\frac{2\pi g_d}{\hbar}
 \sum_{{\bf k},{\bf q}}
 |M_{{\bf k}-{\bf q},{\bf k}}|^2 E_q^2
 \Big[f_{\bf k}(1-f_{{\bf k}-{\bf q}})(1+n_q)
 \nonumber\\
 &\qquad+f_{{\bf k}-{\bf q}}(1-f_{\bf k})n_q\Big]
 \delta(\epsilon_{\bf k}-\epsilon_{{\bf k}-{\bf q}}-E_q).
 \label{eq:zero_kernel}
\end{align}

The corresponding mean heat current, \(Q \equiv \langle \hat{J}\rangle\), is
\begin{align}
 Q &=\frac{2\pi g_d}{\hbar}
 \sum_{{\bf k},{\bf q}}
 |M_{{\bf k}-{\bf q},{\bf k}}|^2 E_q
 \Big[f_{\bf k}(1-f_{{\bf k}-{\bf q}})(1+n_q)
 \nonumber\\
 &\qquad-f_{{\bf k}-{\bf q}}(1-f_{\bf k})n_q\Big]
 \delta(\epsilon_{\bf k}-\epsilon_{{\bf k}-{\bf q}}-E_q).
 \label{eq:mean_current_kernel}
\end{align}
The only structural change between Eqs.~\eqref{eq:zero_kernel} and \eqref{eq:mean_current_kernel} is the sign between the two rates and one additional transferred-energy factor in the noise.

At equilibrium, $T_{ \rm e}=T_{\rm ph}=T$, detailed balance gives $Q=0$.  The linear thermal conductance is
\begin{equation}
 G_{\rm ep}(T)=
 \left.\frac{\partial Q}{\partial T_{ \rm e}}\right|_{T_{\rm e}=T_{\rm ph}=T},
 \label{eq:conductance_def}
\end{equation}
and the zero-frequency thermodynamic FDT is satisfied, \(S(0)=2k_{\rm B} T^2 G_{\rm ep}(T)\). Below we use this relation as a normalization check in every equilibrium limit. At nonzero frequency we do not infer a dynamic thermal conductance from $S(\omega)$.  Finite-frequency heat-current noise can contain extra quantum terms that are not fixed by the static conductance~\cite{Averin2010,Pekola2018}.  All finite-frequency results below are therefore obtained directly from Eq.~\eqref{eq:general_unsym_kernel} and Eq.~\eqref{eq:intro_sym_noise}.

 \section{Zero-frequency results}
\label{sec:zero_results}

This section collects the analytic zero-frequency limits.  The detailed momentum and energy integrations are given in Appendices~\ref{app:bg} and \ref{app:ep}.  Unless stated otherwise, the spin--valley degeneracy $g_d=4$ is included.

\subsection{Bloch--Gr\"uneisen regime}

The BG regime describes low-temperature scattering by
long-wavelength acoustic phonons.  The characteristic momentum available for
intraband scattering is set by twice the Fermi wave vector. Namely, we  consider doped, degenerate graphene, for which
$|\mu|\gg k_{\rm B}T$ and scattering occurs near a
well-defined Fermi surface. The maximum momentum
transfer between states on this Fermi surface is
$2k_{\rm F}$. The corresponding acoustic-phonon energy defines
the BG temperature,
$k_{\rm B}T_{\rm BG}=2\hbar v_{\rm s} k_{\rm F}$.
For $T\ll T_{\rm BG}$, the characteristic thermal phonon
momentum $q_{\rm T}=k_{\rm B}T/(\hbar v_{\rm s})$ is much smaller
than $2k_{\rm F}$, so scattering is dominated by
long-wavelength phonons. For monolayer graphene, the Fermi wave vector and
Bloch--Gr\"uneisen temperature are
\begin{equation}
 k_{\rm F}^{\rm MLG}=\frac{|\mu|}{\hbar v_{\rm F}},
 \qquad
 T_{\rm BG}^{\rm MLG}
 =\frac{2\hbar v_{\rm s} k_{\rm F}^{\rm MLG}}{k_{\rm B}},
 \label{eq:TBG_MLG}
\end{equation}
where \(\mu\) is the chemical potential measured from the charge-neutrality
point, \(v_{\rm F}\) is the monolayer Fermi velocity, and \(v_{\rm s}\) is the
longitudinal sound velocity, the subscript "BG" stands for Bloch--Gr\"uneisen. The use of \(|\mu|\) makes the result applicable
to both electron and hole doping. The typical thermal phonon momentum is of order
\(k_{\rm B}T/(\hbar v_{\rm s})\).  Therefore, the condition
\(T\ll T_{\rm BG}^{\rm MLG}\) implies that the thermally active phonons have
momenta much smaller than \(2k_{\rm F}\).  Scattering is then dominated by
small-momentum processes close to the Fermi surface.  The additional condition
\(|\mu|\gg k_{\rm B}T\) ensures that the electronic distribution is strongly
degenerate and that only a narrow energy window around the Fermi level
contributes. In this regime, the equilibrium zero-frequency heat-current noise and the electron--phonon thermal conductance per unit area are
\begin{align}
 \frac{S_{\rm BG}^{{\rm MLG}}(0)}{A}
 &=
 \frac{8\pi^2}{15}
 \frac{D^2|\mu|k_{\rm B}^5T^5}
 {\rho_{\rm MLG}\hbar^5v_{\rm F}^3v_{\rm s}^3},
 \label{eq:S0_MLG_BG}\\
 \frac{G_{{\rm BG}}^{\rm MLG}(T)}{A}
 &=
 \frac{4\pi^2}{15}
 \frac{D^2|\mu|k_{\rm B}^4T^3}
 {\rho_{\rm MLG}\hbar^5v_{\rm F}^3v_{\rm s}^3},
 \label{eq:G_MLG_BG}
\end{align}
where \(D\) is the deformation-potential coupling and
\(\rho_{\rm MLG}\) is the monolayer areal mass density.  

Equation~\eqref{eq:G_MLG_BG} shows that the BG conductance grows as the third
power of temperature.  This strong suppression at low temperature results
from the small number of thermally occupied acoustic phonons and the limited
electron--phonon scattering phase space.  The conductance is proportional to
\(|\mu|\), mainly because the electronic density of states at the Fermi level
in monolayer graphene increases linearly with the absolute chemical
potential. The noise in Eq.~\eqref{eq:S0_MLG_BG} contains two additional powers of temperature compared with the conductance.  In equilibrium, this follows from
the zero-frequency FDT, in which the noise is
proportional to the conductance multiplied by \(k_{\rm B}T^2\).  Therefore,
a conductance proportional to \(T^3\) produces an equilibrium noise
proportional to \(T^5\).  The finite equilibrium noise reflects continuing
phonon-emission and phonon-absorption events even though their average
contributions to the heat current cancel.

The corresponding mean cooling power agrees with the
Viljas--Heikkil\"a result obtained from the Boltzmann transport equation~\cite{Viljas2010},
\begin{equation}
 \frac{Q_{\rm BG}^{\rm MLG}}{A}
 =
 \Sigma_{\rm MLG}
 \left(T_{\rm e}^4-T_{\rm ph}^4\right),
 \quad
 \Sigma_{\rm MLG}
 =
 \frac{\pi^2D^2|\mu|k_{\rm B}^4}
 {15\rho_{\rm MLG}\hbar^5v_{\rm F}^3v_{\rm s}^3}.
 \label{eq:Q_MLG_BG}
\end{equation}
The coefficient \(\Sigma_{\rm MLG}\) characterizes the strength of the
electron--phonon energy coupling.  When \(T_{\rm e}>T_{\rm ph}\), the expression is
positive and energy flows from electrons to phonons.  When the two
temperatures are equal, the net current vanishes.  Differentiating this
cooling law with respect to the electron temperature at
\(T_{\rm e}=T_{\rm ph}=T\) gives Eq.~\eqref{eq:G_MLG_BG}.

We next consider BLG in the low-energy parabolic-band
approximation.  In this case, the band velocity is not constant but depends on
the Fermi wave vector.  The relevant electronic scales are
\begin{equation}
 k_{\rm F}^{\rm BLG}
 =
 \frac{\sqrt{\gamma_1|\mu|}}{\hbar v_0},
 \qquad
 u_{\rm F}
 =
 2v_0\sqrt{\frac{|\mu|}{\gamma_1}},
 \label{eq:BLG_scales}
\end{equation}
where \(\gamma_1\) is the dominant interlayer hopping energy, \(v_0\) is the
monolayer band velocity entering the effective BLG Hamiltonian in Eq.~(\ref{eq:H_e}), and \(u_{\rm F}\)
is the physical BLG Fermi velocity.  The corresponding
BG temperature is
\(T_{\rm BG}^{\rm BLG}=2\hbar v_{\rm s} k_{\rm F}^{\rm BLG}/k_{\rm B}\).

The parabolic-band description requires the relevant electronic energies to
remain below the interlayer scale, namely
\(|\mu|,k_{\rm B}T\ll\gamma_1\).  The BG result further requires
\(T\ll T_{\rm BG}^{\rm BLG}\) and
\(|\mu|\gg k_{\rm B}T\).  Under these assumptions, the equilibrium BLG noise
and conductance per unit area are
\begin{align}
 \frac{S_{\rm BG}^{{\rm BLG}}(0)}{A}
 &=
 \frac{2\pi^2}{15}
 \frac{D^2\gamma_1 k_{\rm B}^5T^5}
 {\rho_{\rm BLG}\hbar^5v_0^3v_{\rm s}^3}
 \sqrt{\frac{\gamma_1}{|\mu|}},
 \label{eq:S0_BLG_BG}\\
 \frac{G_{{\rm BG}}^{\rm BLG}(T)}{A}
 &=
 \frac{\pi^2}{15}
 \frac{D^2\gamma_1 k_{\rm B}^4T^3}
 {\rho_{\rm BLG}\hbar^5v_0^3v_{\rm s}^3}
 \sqrt{\frac{\gamma_1}{|\mu|}},
 \label{eq:G_BLG_BG}
\end{align}
where \(\rho_{\rm BLG}\) is the bilayer areal mass density.  For an ideal
bilayer it is approximately twice the monolayer areal mass density. The BLG conductance and noise have the same temperature powers as their MLG
counterparts.  The conductance again scales as \(T^3\), while the equilibrium
noise scales as \(T^5\).  These common powers reflect the same two-dimensional
acoustic-phonon phase space and the same low-temperature BG restriction and FDT.

The BLG cooling power in the BG limit is
\begin{equation}
 \frac{Q_{\rm BG}^{\rm BLG}}{A}
 =\Sigma_{\rm BLG}(T_{\rm e}^4-T_{\rm ph}^4), 
 \label{eq:Q_BLG_BG}
\end{equation}
where the cooling-power coefficient is
\begin{equation}
 \Sigma_{\rm BLG}
 =\frac{\pi^2D^2m k_{\rm B}^4}
 {15\rho_{\rm BLG}\hbar^5u_{\rm F}v_{\rm s}^3}.
 \label{eq:Sigma_BLG}
\end{equation}

The important difference is the dependence on carrier density.  In MLG, the
BG coefficient is proportional to \(|\mu|\), whereas in BLG it is
proportional to \(|\mu|^{-1/2}\). The origin of this behavior can be seen from the angular
energy-conservation delta function, whose Jacobian supplies
a factor proportional to the inverse BLG Fermi velocity (see Appendix~B,
Eq.~\eqref{eq:app_mlg_bg_angle}). Because
\(
 u_{\rm F}\propto \sqrt{|\mu|},
\)
this factor behaves as \(1/u_{\rm F}\propto|\mu|^{-1/2}\). The apparent increase of the BLG BG result as \(|\mu|\) approaches zero
should not be interpreted as a physical divergence at charge neutrality.
The derivation assumes a degenerate Fermi surface,
\(|\mu|\gg k_{\rm B}T\), as well as a well-defined BG regime.  Both
assumptions fail sufficiently close to neutrality, where the full
finite-temperature expression must be used.

\subsection{Equipartition regime}
In the EP regime the active phonons are classical, thus, 
\(n_q\simeq k_{\rm B}T/E_q \gg1 \), and the scattering is quasi-elastic. In the EP regime, the thermal conductances of MLG and BLG are obtained using the deformation-potential matrix element in Eq.~\eqref{eq:matrix_element} and the chirality factors in Eqs.~\eqref{eq:chirality_mlg} and~\eqref{eq:chirality_blg}, as derived in Appendix~\ref{app:ep}. The Fermi-window integrals are evaluated exactly within the classical-phonon and quasi-elastic approximations, yielding asymptotic EP expressions whose validity does not extend across the full BG-to-EP crossover. For the two-temperature extension, we assume \(E_q\ll k_{\rm B}T_{\rm ph}\) and \(E_q\ll k_{\rm B}T_{\rm e}\) for the phonons that contribute appreciably to scattering. The resulting equilibrium thermal conductances per unit area are given by
\begin{align}
 & g_{\rm MLG}(\mu,T)=\frac{G_{\rm EP}^{\rm MLG}(T)}{A}=\frac{D^2k_{\rm B}}{30\pi\rho_{\rm MLG}\hbar^5v_{\rm F}^6}\nonumber\\
 &\quad\times\Big[15\mu^4+30\pi^2\mu^2(k_{\rm B}T)^2
 +7\pi^4(k_{\rm B}T)^4\Big]
 \label{eq:G_MLG_EP}
 \end{align}
 
\begin{align}
 g_{\rm BLG}(\mu,T)
 &=\frac{G_{\rm EP}^{\rm BLG}(T)}{A}
 =\frac{D^2\gamma_1^3k_{\rm B}^2T}
 {2\pi\rho_{\rm BLG}\hbar^5v_0^6}
 \nonumber\\
 &\quad\times
 \ln\!\left[2\cosh\left(\frac{\mu}{2k_{\rm B}T}\right)\right].
 \label{eq:G_BLG_EP}
\end{align}
In Eq.~(\ref{eq:G_BLG_EP}) above, the sound velocity cancels in the EP conductance because the classical Bose factor contributes $1/(v_{\rm s}q)$, while the matrix element and transferred energy supply the compensating factors. One can again straightforwardly show that the zero-frequency noise again satisfies the FDT, namely \(S^l_{\rm EP}(0) = 2k_{\rm B}T^2 G^l_{\rm EP}(T)\), where \(l={\rm MLG, BLG}\).
\begin{table*}[t]
 \caption{Zero-frequency equilibrium power laws and conductances for physical MLG and BLG.  All quantities are per area. See Eqs.~\eqref{eq:Q_MLG_BG},~\eqref{eq:Sigma_BLG},  \eqref{eq:G_MLG_EP} and \eqref{eq:G_BLG_EP}}
 \label{tab:zero_results}
 \begin{ruledtabular}
 \begin{tabular}{lccc}
 System and Regime & Cooling law & $G_{\rm ep}/A$ & $S(0)/A$ \\
 \hline
 MLG BG & $\Sigma_{\rm MLG}(T_{\rm e}^4-T_{\rm ph}^4)$ & $4\Sigma_{\rm MLG}T^3$ & $8k_{\rm B}\Sigma_{\rm MLG}T^5$ \\
 MLG EP & $g_{\rm MLG}(\mu,T_{\rm e})(T_{\rm e}-T_{\rm ph})$ & $g_{\rm MLG}(\mu,T)$ & $2k_{\rm B}T^2g_{\rm MLG}(\mu,T)$ \\
 BLG BG & $\Sigma_{\rm BLG}(T_{\rm e}^4-T_{\rm ph}^4)$ & $4\Sigma_{\rm BLG}T^3$ & $8k_{\rm B}\Sigma_{\rm BLG}T^5$ \\
 BLG EP & $g_{\rm BLG}(\mu,T_{\rm e})(T_{\rm e}-T_{\rm ph})$ & $g_{\rm BLG}(\mu,T)$ & $2k_{\rm B}T^2g_{\rm BLG}(\mu,T)$ \\
 \end{tabular}
 \end{ruledtabular}
\end{table*}

\subsection{Two-temperature regime}

We now consider electrons and phonons maintained at separate
temperatures, $T_{\rm e}$ and $T_{\rm ph}$. In the BG regime,
both temperatures are assumed to satisfy
$T_{\rm e},T_{\rm ph}\ll T_{\rm BG}$, while the electrons remain
degenerate, $k_{\rm B}T_{\rm e}\ll|\mu|$. Within the degenerate Fermi-surface reduction described in
Appendix~\ref{app:bg}, the electronic energy integral is extended
to the full real axis. The zero-frequency noise is obtained from the sum of the
emission and absorption terms in Eq.~\eqref{eq:zero_kernel}. Next, using the continuum and angular reductions in the intgerations with respect to energy variable (see 
Appendix~\ref{app:bg}), the MLG noise at zero frequency becomes
\begin{align}
 \frac{S_{\rm BG}^{\rm MLG}
 (0;T_{\rm e},T_{\rm ph})}{A}
 &=\frac{g_dD^2|\mu|}
 {4\pi^2\rho_{\rm MLG}\hbar^5v_{\rm F}^3v_{\rm s}^3}
 \nonumber\\
 &\times
 \int_0^\infty dE\,E^4
 \left[n_{\rm e}+n_{\rm ph}
       +2n_{\rm e}n_{\rm ph}\right],
 \label{eq:noneq_BG_MLG}
\end{align}
where $g_d=4$ is the physical spin--valley degeneracy, $n_{\rm ph}$ is the phonon occupation, whereas $n_{\rm e}$
is an auxiliary Bose function generated by the electron
phase-space integrals,
\begin{equation}
 n_i(E)=\frac{1}{\exp[E/(k_{\rm B}T_i)]-1},
 \qquad i={\rm e},{\rm ph}.
 \label{eq:noneq_bose_functions}
\end{equation}
The BLG result contains the same energy integral, with
the diffrent prefactor
\begin{align}
 \frac{S_{\rm BG}^{\rm BLG}
 (0;T_{\rm e},T_{\rm ph})}{A}
 &=\frac{g_dD^2\gamma_1}
 {16\pi^2\rho_{\rm BLG}\hbar^5v_0^3v_{\rm s}^3}
 \sqrt{\frac{\gamma_1}{|\mu|}}
 \nonumber\\
 &\times
 \int_0^\infty dE\,E^4
 \left[n_{\rm e}+n_{\rm ph}
       +2n_{\rm e}n_{\rm ph}\right].
 \label{eq:noneq_BG_BLG_prefactor}
\end{align}
Thus the two temperatures enter separately through the
electronic phase space and the phonon occupation. Prefactors in Eqs.~(\ref{eq:noneq_BG_MLG}) and (\ref{eq:noneq_BG_BLG_prefactor}) are consistent with the finite-frequency
coefficients in Eqs.~\eqref{eq:C_BG_MLG}
and~\eqref{eq:C_BG_BLG} of next section.

We further parameterize the temperature imbalance by \(T_{\rm e/ph}=T \pm \Delta T/2\) and thus the noise is even in $\Delta T$.
For $|\Delta T|\ll T$, expansion at fixed $T$ and $\mu$
gives, for both MLG and BLG,
\begin{equation}
\frac{S_{\rm BG}^{l}(0;T_{\rm e},T_{\rm ph})}
 {S_{\rm BG}^{l}(0;T,T)}=1+\frac{5\pi^2}{21}
       \left(\frac{\Delta T}{T}\right)^2,
 \quad l={\rm MLG},{\rm BLG}.
 \label{eq:noneq_BG_expansion}
\end{equation}
The sub-leadning correction is propotional to \( \propto (\Delta T/T)^4\). In the EP regime, we instead use the classical-phonon
and quasi-elastic reduction of Appendix~\ref{app:ep}.
The phonon occupation is evaluated at $T_{\rm ph}$,
while the electronic Fermi window is evaluated at
$T_{\rm e}$, i.e. \(n_{\rm ph}(E_q)\simeq
 k_{\rm B}T_{\rm ph}/E_q\) and \( f_{\rm e}(1-f_{\rm e})
 =-k_{\rm B}T_{\rm e}
 (\partial f_{\rm e}/\partial\epsilon)\) and consequently 
The microscopic emission-plus-absorption rate sum
then yields
\begin{equation}
 \frac{S_{\rm EP}^{l}(0;T_{\rm e},T_{\rm ph})}{A}
 =2k_{\rm B}T_{\rm e}T_{\rm ph}\,
 g_l(\mu,T_{\rm e}),
 \label{eq:noneq_EP_common}
\end{equation}
where \(g_l\) are given in Eqs.~\eqref{eq:G_MLG_EP}
and~\eqref{eq:G_BLG_EP}, respectively.
The units of $g_l$ are
$\mathrm{W\,m^{-2}\,K^{-1}}$. This expression is derived directly from the rate sum.
At $T_{\rm e}=T_{\rm ph}=T$, it recovers the equilibrium
identity $S_{\rm EP}^{l}(0)=2k_{\rm B}T^2G_{\rm EP}^{l}(T)$.
Away from equilibrium, a thermal response must be defined
from the mean heat current in
Eq.~\eqref{eq:mean_current_kernel}, with the held-fixed
variables specified; it cannot be inferred from the
noise alone.

\section{Finite-frequency spectra}
\label{sec:finite_results}

Finite frequency introduces the new energy scale  $\hbar\omega$.  The emission and absorption vertices in Eq.~\eqref{eq:general_unsym_kernel} become $\hbar \omega \mp 2E_q$ after the delta functions are imposed.  The symmetrized spectrum is even in $\omega$, but its frequency dependence is not fixed by the zero-frequency conductance.  We now give the reduced BG and EP spectra used in the numerical analysis.

\subsection{Bloch--Gr\"uneisen spectrum}
In doped MLG the BG reduction keeps the intraband Fermi surface, linearizes the electron energy difference, and retains the full transferred phonon energy inside the Fermi functions.  In this case, using the Eq.~(\ref{eq:intro_sym_noise}) one obtains the following equation for the finite-frequency noise
\begin{equation}
\frac{S^l_{\rm BG}(\omega)}{A}
 =\frac{C^l_{\rm BG}}{2}\int_0^{\infty}dE\,E
 \left[\mathcal B(E, \hbar \omega)+\mathcal B(E,-\hbar \omega)\right],
 \label{eq:BG_sym_reduced}
\end{equation}
where \(l={\rm MLG, BLG}\) and the frequency dependent part of integrand is given by 
\begin{equation}
\begin{split}
 & {\cal B}(E, \hbar \omega)=(\hbar \omega-2E)^2[1+n_{\rm ph}(E,T_{\rm ph})]\Phi_{\rm e}(E-\hbar \omega;T_{\rm e}) \\
 & +(\hbar \omega+2E)^2n_{\rm ph}(E,T_{\rm ph})\Psi_{\rm e}(E+\hbar \omega;T_{\rm e}).
 \label{eq:Frequecny dependent factor beta(E,omega)}
\end{split}
\end{equation}
In the above Eq.~(\ref{eq:Frequecny dependent factor beta(E,omega)}) The thermal transition factors are related by the symmetry relation \( \Phi_{\rm e}(x;T_{\rm e})= \Psi_{\rm e}(-x;T_{\rm e})\) and \(
 \Phi_{\rm e}(x;T_{\rm e})= 
\int d\epsilon\,
 f_{\rm e}(\epsilon+x;T_{\rm e})[1-f_{\rm e}(\epsilon;T_{\rm e})]=x/[\exp(x/k_{\rm B}T_{\rm e})-1]\). The prefactors in front of the integral which correspond to MLG and BLG  have the form   
\begin{equation}
 C_{\rm BG}^{\rm MLG}
 =\frac{g_dD^2|\mu|}{16\pi^2\rho_{\rm MLG}\hbar^5v_{\rm F}^3v_{\rm s}^3},
 \label{eq:C_BG_MLG}
\end{equation}
\begin{equation}
 C_{\rm BG}^{\rm BLG}
 =\frac{g_dD^2\gamma_1}{64\pi^2\rho_{\rm BLG}\hbar^5v_0^3v_{\rm s}^3}
 \sqrt{\frac{\gamma_1}{|\mu|}}.
 \label{eq:C_BG_BLG}
\end{equation}
Eq.~(\ref{eq:BG_sym_reduced}) expression is evaluated without expanding in $\hbar \omega$ in the ``exact reduced'' BG curves below.

At equilibrium and $|\hbar \omega|\ll k_{\rm B}T$, Eq.~\eqref{eq:BG_sym_reduced} gives
\begin{equation}
 \frac{S_{\rm BG}(\omega;T,T)}
 {S_{\rm BG}(0;T,T)}
 =1+\left(\frac{1}{12}+\frac{5}{16\pi^2}\right)
 \left(\frac{\hbar\omega}{k_{\rm B}T}\right)^2,
 \label{eq:BG_quadratic}
\end{equation}
and the sub-leading correction is proportional to \((\hbar \omega/k_{\rm B}T)^4\). In the opposite quantum limit $T_{\rm e}=T_{\rm ph}=T=0$, the same BG kernel gives
\begin{equation}
 \frac{S_{\rm BG}^l(\omega;0,0)}{A}
 =\frac{C_{\rm BG}^l}{60}|\hbar\omega|^5,
 \qquad |\hbar\omega|\ll k_{\rm B}T_{\rm BG},
 \label{eq:BG_zero_temperature}
\end{equation}
showing explicitly that finite-frequency heat-current noise survives when the zero-frequency thermal noise vanishes.  This is the graphene analogue of the quantum heat-noise effect discussed in Ref.~\cite{IdrisovAnyonicHeat2024}.

For the symmetric temperature bias , introducing the dimensionless variables \(r=\Delta T/T\) , \(w=\hbar \omega/k_{\rm B} T\) and \(x=E/k_{\rm B}T\), the Eq.~\eqref{eq:BG_sym_reduced} results in
\begin{align}
 \frac{S_{\rm BG}^l(\omega;T_{\rm e},T_{\rm ph})}{A}
 &=\frac{C^l_{\rm BG}}{2}(k_{\rm B}T)^5\nonumber\\
 &\quad\times\int_0^{\infty}dx\,x
 \left[\mathcal B_r(x,w)+\mathcal B_r(x,-w)\right].
 \label{eq:BG_dimensionless}
\end{align}
with the explicit form of funtion, \(\mathcal B_r(x,w)\), are given in Appendix~\ref{app:finite}.  Eq.~\eqref{eq:BG_dimensionless} is the basis of the nonequilibrium BG ratio and heat-map calculations.

\subsection{Equipartition spectrum}
We obtain the finite-frequency EP spectra by applying
the classical-phonon and quasi-elastic approximations
to the general finite-frequency kernel in
Eq.~\eqref{eq:general_unsym_kernel}.
For the phonon modes contributing to the integral, \(n_{\rm ph}(E_q,T_{\rm ph})
 \simeq k_{\rm B}T_{\rm ph}/E_q\gg1\), thus \( 1+n_{\rm ph}\simeq n_{\rm ph}\). In the energy-conservation delta functions, the phonon
energy is neglected relative to the electronic
energy scales, i.e. \(\delta(\hbar\omega-\Delta\mp E_q)
 \simeq \delta(\hbar\omega-\Delta)\), where $\Delta$ denotes the electronic energy difference from Eq.~(\ref{eq:delta_energy}).
Thus, at finite frequency the electronic states remain
separated by the probe energy $\hbar\omega$.
The phonon energy must nevertheless be retained in
the two current vertices. Using these approximations in the two shell
contributions of Eq.~\eqref{eq:general_unsym_kernel},
the classical-phonon vertex sum becomes \( n_{\rm ph}
 (\hbar\omega-2E_q)^2+n_{\rm ph}
 (\hbar\omega+2E_q)^2 \simeq (2k_{\rm B}T_{\rm ph}/E_q
 )\left[(\hbar\omega)^2+4E_q^2\right]\). The terms linear in $\hbar\omega E_q$ cancel between
the two vertices. It is worth pointing out, that the term proportional to
$(\hbar\omega)^2$ vanishes at zero frequency and
produces an additional dynamical
contribution in case of finite frequency case, under the consideration. For MLG, substituting the above classical-phonon vertex sum into Eq.~\eqref{eq:general_unsym_kernel},
performing the angular integrals, and symmetrizing
in frequency gives
\begin{equation}
\begin{split}
 & \frac{S_{\rm EP}^{\rm MLG}
 (\omega;T_{\rm e},T_{\rm ph})}{A}=\frac{D^2k_{\rm B}T_{\rm ph}}
 {2\pi\rho_{\rm MLG}\hbar^5v^2_{\rm s}v_{\rm F}^4} \\ & \times \left[(\hbar\omega/2)^2 \mathcal{I}_2(|\hbar\omega|;T_{\rm e})+(v_s/v_{\rm F})^2\mathcal{I}_4(|\hbar\omega|;T_{\rm e})\right].
 \label{eq:EP_MLG_finite}
\end{split}
\end{equation}
The angular reduction leading to this expression
is detailed in Appendix~\ref{app:finite}. The electronic occupation factor entering the
symmetrized spectrum is \( K(\epsilon;\omega;T_{\rm e})=f_{\rm e}(\epsilon;T_{\rm e})
 [1-f_{\rm e}(\epsilon+\hbar\omega;T_{\rm e})]+
 f_{\rm e}(\epsilon+\hbar\omega;T_{\rm e})
 [1-f_{\rm e}(\epsilon;T_{\rm e})] \), where
\(f_{\rm e}(\epsilon;T_{\rm e}) \) is the FD distribution function. This factor adds the occupation
probabilities for transitions between electronic
states of energies $\epsilon$ and
$\epsilon+\hbar\omega$ in both directions.
It follows from symmetrizing the electronic rate
factors in Eq.~\eqref{eq:general_unsym_kernel}. The two electronic integrals appearing in
Eq.~\eqref{eq:EP_MLG_finite} are
\begin{equation}
\begin{split}
 & \mathcal{I}_2(\hbar\omega;T_{\rm e})
 =\int d\epsilon\,
 |\epsilon||\epsilon+\hbar\omega| \Theta[\epsilon(\epsilon+\hbar\omega)]
 K(\epsilon;\omega;T_{\rm e}), \\
 & \mathcal{I}_4(\hbar\omega;T_{\rm e})
=\int d\epsilon\,
 |\epsilon||\epsilon+\hbar\omega| \Theta[\epsilon(\epsilon+\hbar\omega)] K(\epsilon;\omega;T_{\rm e}) \\
 &\quad\times
 \left[
 (\epsilon+\hbar\omega)^2+\epsilon^2
 -|\epsilon||\epsilon+\hbar\omega|
 \right]. 
\end{split}
\label{eq:I2andI4}
\end{equation} In Eq.~\eqref{eq:I2andI4}, the factors
$|\epsilon||\epsilon+\hbar\omega|$ arise from
the two MLG radial density-of-states measures.
The additional polynomial in Eq.~\eqref{eq:I2andI4}
comes from the angular average of the
momentum-dependent vertex and chirality factor.
Consequently, \(\mathcal{I}_2\) weights the
\((\hbar\omega)^2\) contribution in classical-phonon vertex sum, whereas \(\mathcal{I}_4\)
weights its \(4E_q^2\) contribution.
The step function restricts both integrals to
same-band intraband transitions.
Both integrals are even in $\omega$, which
allows their arguments in
Eq.~\eqref{eq:EP_MLG_finite} to be written
as \(|\hbar\omega|\).

For BLG, the same reduction of
Eq.~\eqref{eq:general_unsym_kernel} uses the
parabolic dispersion and BLG chirality factor,
as detailed in Appendix~\ref{app:finite}.
The resulting spectrum is
\begin{equation}
\begin{split}
& \frac{S_{\rm EP}^{\rm BLG}
 (\omega;T_{\rm e},T_{\rm ph})}{A}=\frac{D^2\gamma_1^2k_{\rm B}T_{\rm ph}}
 {8\pi\rho_{\rm BLG}\hbar^5v^2_{\rm s}v_0^4} \\
& 
 \times [(\hbar\omega/2)^2 \mathcal{J}_0(|\hbar\omega|;T_{\rm e})+(\gamma_1 v^2_{\rm s}/v^2_0)\mathcal{J}_1(|\hbar\omega|;T_{\rm e})].
 \label{eq:EP_BLG_finite}
\end{split}
\end{equation}
The electronic integrals in this expression are
\begin{equation}
\begin{split}
& \mathcal{J}_0(\hbar\omega;T_{\rm e})
=\int d\epsilon\,
 K(\epsilon;\omega;T_{\rm e})
 \Theta[\epsilon(\epsilon+\hbar\omega)],\\
& \mathcal{J}_1(\hbar\omega;T_{\rm e})=\int d\epsilon
 [|\epsilon|+|\epsilon+\hbar\omega|] K(\epsilon;\omega;T_{\rm e})
 \Theta[\epsilon(\epsilon+\hbar\omega)].
\end{split}
 \label{eq:J1andJ0}
\end{equation}
The constant radial density of states of the
parabolic BLG bands gives the unweighted
phase-space integral \(\mathcal{J}_0\) in Eq.~\eqref{eq:J1andJ0}.
The angular average of the momentum-dependent
vertex introduces the energy weight
\(|\epsilon|+|\epsilon+\hbar\omega|\) in
\( \mathcal{J}_1\). Thus, \(\mathcal{J}_0\) and \(\mathcal{J}_1\) describe, respectively,
the \((\hbar\omega)^2\) and \(4E_q^2\) terms of classical vertex sum. The step functions again correspond to the same-band
intraband scattering.

The reduced EP curves are obtained by numerical
evaluation of Eqs.~\eqref{eq:EP_MLG_finite}
and \eqref{eq:EP_BLG_finite} within the
classical-phonon and quasi-elastic approximations. The low-frequency behavior of heat current noise for MLG in EP regime in the leading order of finite frequencies has quadratic form
\begin{equation}
 \frac{S_{\rm EP}^{\rm MLG}(\omega)}
 {S_{\rm EP}^{\rm MLG}(0)}=1+(\hbar\omega)^2
 \left[
 \frac{1}{12(k_{\rm B}T_{\rm e})^2}+\frac{M_2}{M_4}
 \left(1+\frac{v_{\rm F}^2}{4v_{\rm s}^2}\right)
 \right],
\label{eq:EP_MLG_quadratic}
\end{equation}
where the Fermi-window moments follow from
$M_j=\int_{-\infty}^{\infty}d\epsilon\,
\epsilon^j(-\partial f_{\rm e}/\partial\epsilon)$
for $j=2,4$ and have the form
\begin{align}
 M_2(\mu,T_{\rm e})
 &=\mu^2+\frac{\pi^2}{3}(k_{\rm B}T_{\rm e})^2,\\
 M_4(\mu,T_{\rm e})
 &=\mu^4+2\pi^2\mu^2(k_{\rm B}T_{\rm e})^2
 +\frac{7\pi^4}{15}(k_{\rm B}T_{\rm e})^4.
\end{align}
In the degenerate BLG EP limit, the low-frequency dependence is given by 
\begin{equation}
 \frac{S_{\rm EP}^{\rm BLG}(\omega)}
 {S_{\rm EP}^{\rm BLG}(0)}=1+(\hbar \omega)^2 \left[\frac{1}{12(k_{\rm B}T_{\rm e})^2}
 +\frac{1}{16m v_{\rm s}^2|\mu|}\right].
 \label{eq:Noise BLG EP low frequency}
\end{equation}
The sub-leading corrections to Eq.~(\ref{eq:EP_MLG_quadratic}) and Eq.~(\ref{eq:Noise BLG EP low frequency}) proportional to \(\propto (\hbar \omega)^4\). The detailed calculations are provided in Appendix~\ref{app:finite}.

\subsection{Full on-shell MLG benchmark}
To connect the two limiting theories, we also evaluate
the doped intraband MLG spectrum without the BG
small-$q$ or EP classical-phonon reductions.
Momentum conservation gives
${\bf p}={\bf k}+{\bf q}$.
The angular integration over the energy-conservation
delta function is performed analytically
(see Appendix~\ref{app:numerics}).
The resulting shell variables are 
\begin{align}
 p_{\lambda s}(k,q)&=k+s\frac{v_{\rm s}}{v_{\rm F}}q-\lambda\frac{\omega}{v_{\rm F}},
 \label{eq:full_p}\\
 \mu_{\lambda s}(k,q)&=\frac{p_{\lambda s}^2-k^2-q^2}{2kq},
 \label{eq:full_mu}
\end{align}
where $\lambda=\pm1$ labels the two frequency signs and $s=+1$ ($-1$) labels emission (absorption).  The scattering event is allowed only if $p_{\lambda s}>0$ and $|\mu_{\lambda s}|<1$.  The exact chirality factor is
\begin{equation}
 F_{++}=\frac{1+\cos\phi}{2},
 \qquad
 \cos\phi=\frac{k+q\mu_{\lambda s}}{p_{\lambda s}}.
 \label{eq:full_chirality}
\end{equation}
The resulting two-dimensional on-shell integral is written in Appendix~\ref{app:numerics}.  It retains the exact Bose and Fermi functions and the vertex $(\lambda\hbar\omega-2sE_q)^2$.  We refer to this result as the ``full on-shell'' intraband calculation. This numerical calculation retains the exact intraband
scattering geometry and distribution functions of the chosen doped MLG model. Its low- and
high-temperature limits give the BG and EP reductions.

\section{Numerical results}
\label{sec:numerics}

We now evaluate the analytic formulas and the full doped-intraband MLG integral.  The purpose of the numerical analysis is threefold.  First, it verifies the zero-frequency FDT normalization and the Viljas--Heikkil\"a limiting laws~\cite{Viljas2010}.  Second, it determines the range over which the low-frequency expansions describe the full reduced spectra.  Third, it shows how the non-equilibrium temperature bias changes the finite-frequency correction.

Unless noted otherwise, we use the parameters listed in
Table~\ref{tab:parameters}. The chosen deformation-potential value \(D\) is
representative rather than universal; all conductances and heat-current noise
spectra scale quadratically with \(D\). For BLG, we take the areal mass density
to be twice the monolayer value. The figures are presented in terms of the
ordinary frequency \(f\), while the analytical expressions use the angular
frequency \(\omega\). The corresponding probe energy may therefore be written
equivalently as \(\hbar\omega\) or \(hf\), with \(\omega=2\pi f\). The numerical curves are evaluated from the zero-frequency
BG and EP expressions in 
Eqs.~\eqref{eq:S0_MLG_BG}, \eqref{eq:G_MLG_BG},
\eqref{eq:S0_BLG_BG}, \eqref{eq:G_BLG_BG},
\eqref{eq:G_MLG_EP}, and \eqref{eq:G_BLG_EP},
and from the reduced finite-frequency spectra in
Eqs.~(37), (51), and (54).
The full doped-intraband MLG benchmark is evaluated
from Eq.~(E6), while the quadratic approximations
follow Eqs.~(43), (55), and (D18) given in
Secs.~\ref{sec:zero_results} and \ref{sec:finite_results}.

\begin{table}[t]
 \caption{Parameters used in the numerical figures.
Material parameters follow representative graphene
values discussed in Refs.~\cite{Viljas2010,McCann2013}.}
 \label{tab:parameters}
 \begin{ruledtabular}
 \begin{tabular}{lc}
 quantity & value \\
 \hline
 deformation potential $D$ & $20\ \mathrm{eV}$ \\
 chemical potential $\mu$ & $0.10\ \mathrm{eV}$ \\
 band velocity $v_{\rm F}=v_0$ & $1.0\times10^6\ \mathrm{m\,s^{-1}}$ \\
 sound velocity $v_{\rm s}$ & $2.0\times10^4\ \mathrm{m\,s^{-1}}$ \\
 interlayer coupling $\gamma_1$ & $0.39\ \mathrm{eV}$ \\
 MLG mass density $\rho_{\rm MLG}$ & $7.6\times10^{-7}\ \mathrm{kg\,m^{-2}}$ \\
 BLG mass density $\rho_{\rm BLG}$ & $1.52\times10^{-6}\ \mathrm{kg\,m^{-2}}$ \\
 spin--valley degeneracy $g_d$ & $4$
 \end{tabular}
 \end{ruledtabular}
\end{table}

\begin{figure*}[t]
 \centering
 \includegraphics[width=\textwidth]{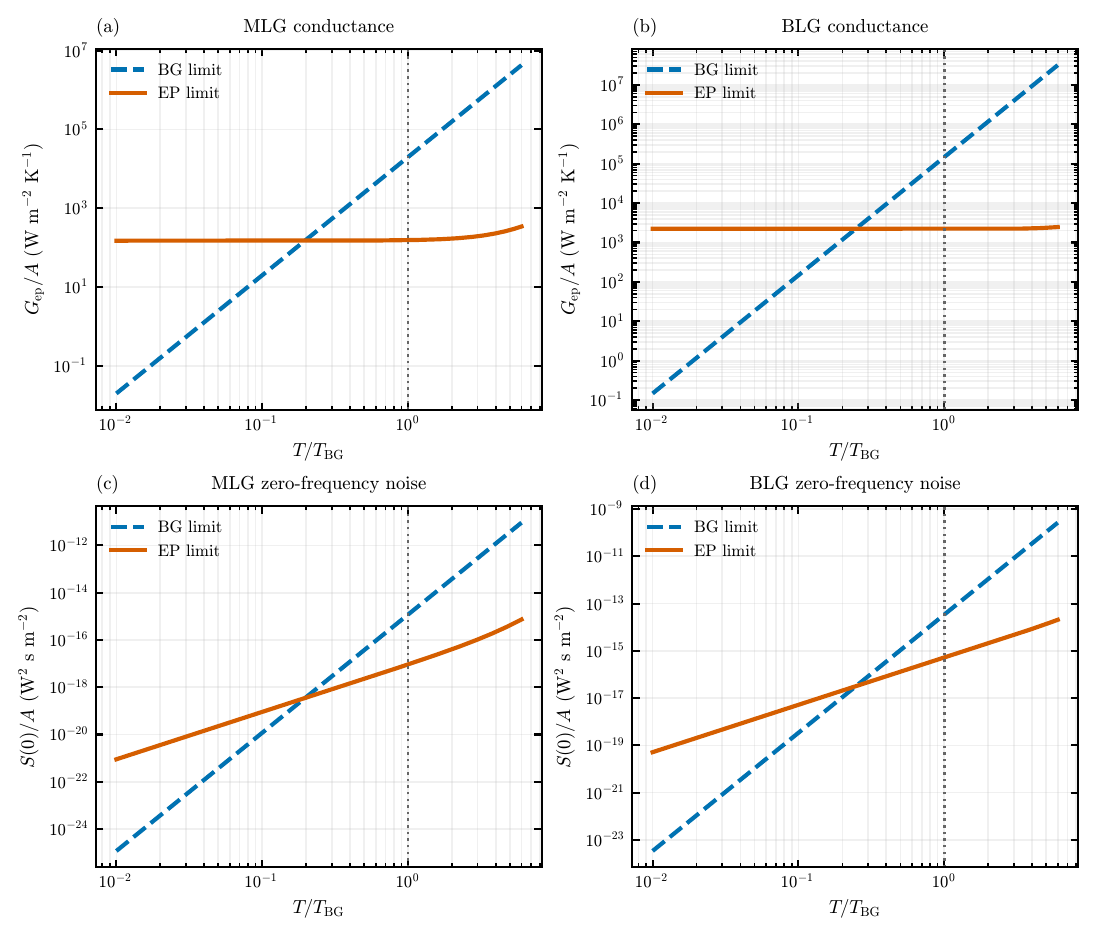}
 \caption{Equilibrium thermal conductance and zero-frequency heat-current noise in the BG and EP limits.
 The BG conductances follow
Eqs.~\eqref{eq:G_MLG_BG} and \eqref{eq:G_BLG_BG};
the EP conductances follow
Eqs.~\eqref{eq:G_MLG_EP} and \eqref{eq:G_BLG_EP}.
The temperature is normalized by the appropriate BG temperature of MLG or BLG.  Dashed curves show the BG laws and solid curves the EP laws.  The lower panels follow from the upper panels through $S(0)=2k_{\rm B}T^2G_{\rm ep}(T)$.}
 \label{fig:viljas_fdt}
\end{figure*}

Fig.~\ref{fig:viljas_fdt} shows the equilibrium linear
electron--phonon thermal conductance per unit area,
$G_{\rm ep}/A$, for both MLG and BLG in the BG and EP
limits. Here
$G_{\rm ep}(T)=
\left.\partial Q/\partial T_{\rm e}\right|_{T_{\rm e}=T_{\rm ph}=T}$,
as defined in Eq.~(\ref{eq:conductance_def}).
The corresponding zero-frequency noise curves are
constructed using the equilibrium relation
$S(0)/A=2k_{\rm B}T^2[G_{\rm ep}(T)/A]$.
The conductance expressions are asymptotic limits;
their extrapolations outside the respective BG and EP
regimes do not describe the full crossover.These curves reproduce the first-moment limiting laws of Ref.~\cite{Viljas2010} and provide the equilibrium normalization of the second moment.


\begin{figure*}[t]
 \centering
 \includegraphics[width=\textwidth]{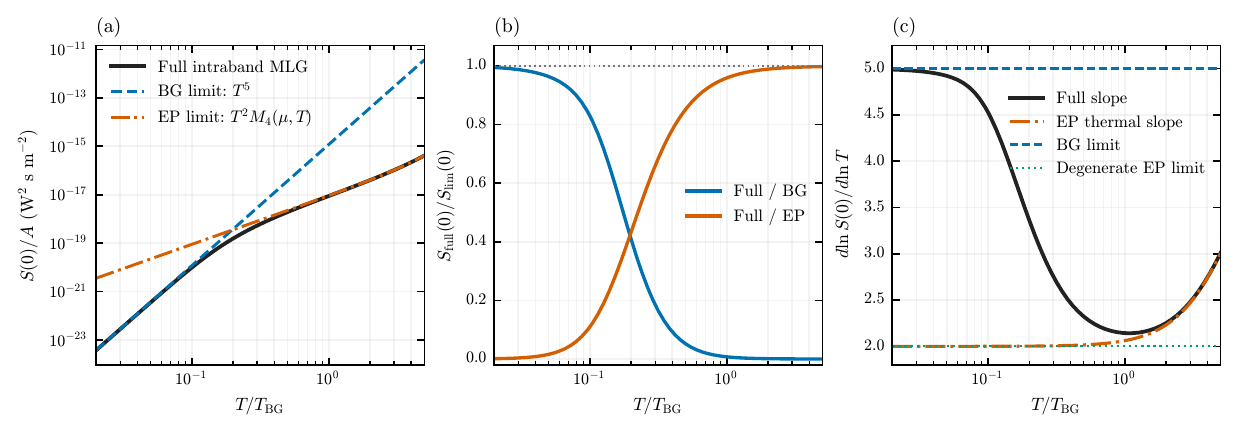}
 \caption{Full zero-frequency MLG crossover.  (a) Full on-shell noise compared with BG and EP limits obtained from
Eq.~\eqref{eq:app_full_integral} at $\omega=0$,
compared with the BG expression,
Eq.~\eqref{eq:S0_MLG_BG}, and the EP noise
$S_{\rm EP}(0)/A=2k_{\rm B}T^2g_{\rm MLG}(\mu,T)$,
with $g_{\rm MLG}$ given by
Eq.~\eqref{eq:G_MLG_EP}.The BG and EP limits correspond to $T\ll T_{\rm BG}$
and $T\gg T_{\rm BG}$, respectively.
The BG noise scales as $T^5$, whereas the MLG EP
noise is proportional to $T^2M_4(\mu,T)$, with
$M_4$ defined in Eq.~(61).
The latter reduces to $T^2$ for
$k_{\rm B}T\ll|\mu|$.
Limiting curves outside their respective regimes
are extrapolations. (b) Ratios $S_{\rm full}(0)/S_{\rm lim}(0)$,
where $S_{\rm lim}(0)$ denotes the BG or EP
limiting expression.  (c) Local logarithmic slope $d\ln S /d\ln T$.  The full result approaches the BG law at low temperature and the EP law at high temperature.}
 \label{fig:zero_crossover}
\end{figure*}

The limiting formulas do not by themselves describe the crossover.  Fig.~\ref{fig:zero_crossover} therefore evaluates the full zero-frequency MLG on-shell integral.  The low-temperature numerical curve approaches the $T^5$ BG law, whereas the high-temperature curve approaches the EP result.  The ratio panel gives a direct validity check, and the local logarithmic slope evolves from approximately $5$ toward the degenerate EP value $2$.  The crossover is smooth and occurs over a broad interval around $T_{\rm BG}$, rather than at a sharp boundary.

\begin{figure*}[t]
 \centering
 \includegraphics[width=\textwidth]{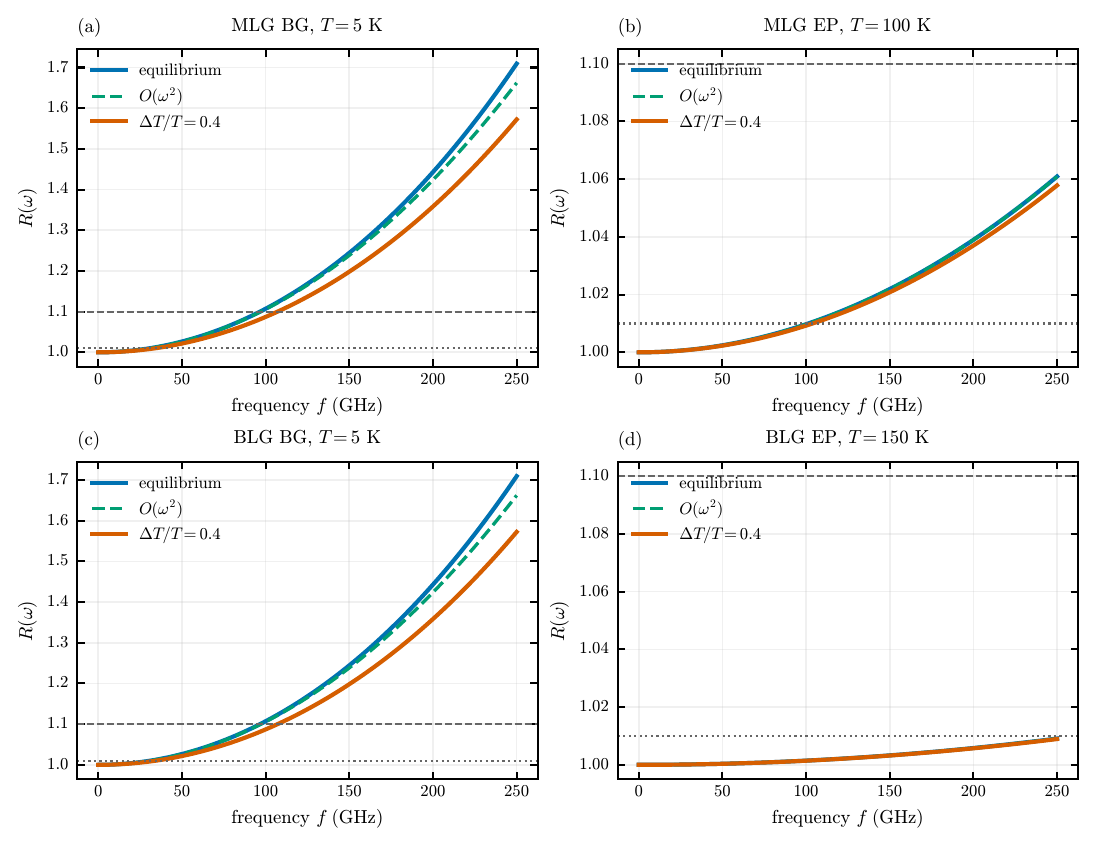}
 \caption{Finite-frequency noise ratio $R=S(\omega)/S(0)$ in the reduced BG and EP theories.  The equilibrium curves use $T_{\rm e}=T_{\rm ph}$; the nonequilibrium curves use the symmetric bias with $\Delta T/T=0.4$.  Dashed lines show the quadratic low-frequency approximations.  Horizontal guides mark $1\%$ and $10\%$ corrections.}
 \label{fig:finite_ratios}
\end{figure*}

In order to investigate the exact curves it is practical to introduce the dimensionless observable \cite{Pekola2018}
\begin{equation}
 R(\omega;T_{\rm e},T_{\rm ph})=
 \frac{S(\omega;T_{\rm e},T_{\rm ph})}
 {S(0;T_{\rm e},T_{\rm ph})}.
 \label{eq:R_definition}
\end{equation}
It removes the overall $D^2$ and area dependence and highlights the dynamical part of the fluctuations.  Figure~\ref{fig:finite_ratios} shows $R$ for the exact reduced BG and EP spectra.  The equilibrium and nonequilibrium curves are displayed together with the quadratic low-frequency approximation.  At small frequency the exact and asymptotic curves coincide.  Their separation at higher frequency is caused by the $O(\omega^4)$ and higher terms omitted from the expansion.

The BG ratio is the same for MLG and BLG within the leading Fermi-surface reduction because their different material factors cancel in Eq.~\eqref{eq:R_definition}.  The EP ratios differ because the two electronic dispersions give different energy moments.  For the parameters of Table~\ref{tab:parameters}, finite-frequency corrections are negligible at ordinary low-GHz frequencies, but reach the percent level in the high-GHz to sub-THz window. The characteristic frequency scale depends on the
electronic and phononic temperature scales.
In the BG regime, the normalized spectrum depends
on both $T_{\rm e}$ and $T_{\rm ph}$ through the
electronic phase-space and phonon-occupation factors
in Eq.~(38).
Within the classical-phonon EP reduction,
Eqs.~(51) and (54) contain an overall factor
$T_{\rm ph}$, which cancels in
$R(\omega)=S(\omega)/S(0)$.
The normalized EP frequency dependence therefore
depends on $hf/(k_{\rm B}T_{\rm e})$, the chemical
potential, and the material parameters, while the
absolute noise retains its dependence on $T_{\rm ph}$.

\begin{figure*}[t]
 \centering
 \includegraphics[width=\textwidth]{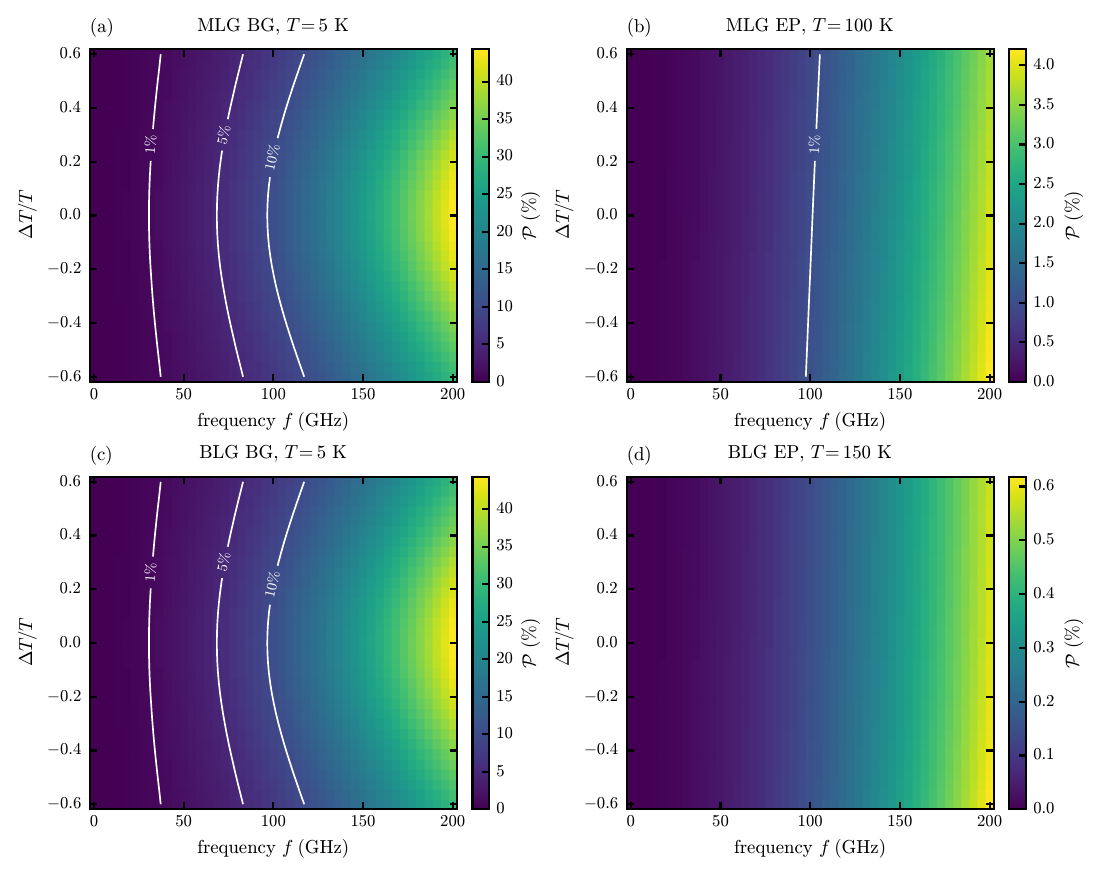}
 \caption{Nonequilibrium finite-frequency percentage correction ${\cal P}$ from Eq.~\eqref{eq:percentage_definition}.  The horizontal axis is ordinary frequency and the vertical axis is the symmetric temperature bias.  Contours identify $1\%$, $5\%$, and $10\%$ corrections.  The maps demonstrate that finite-frequency noise outside equilibrium must be evaluated directly rather than inferred from a conductance.}
 \label{fig:heatmaps}
\end{figure*}

The non-equilibrium dependence is summarized by the percentage correction~\cite{Pekola2018}
\begin{equation}
 {\cal P}(f,\Delta T/T)=100\left[
 R(\omega;T_{\rm e},T_{\rm ph})-1\right],
 \label{eq:percentage_definition}
\end{equation}
with symetric temperature bias $T_{\rm e/ph}=T\pm \Delta T/2$.  Fig.~\ref{fig:heatmaps} plots this quantity for MLG and BLG in both limiting regimes.  The BG maps are obtained from Eq.~\eqref{eq:BG_sym_reduced}; the EP maps use Eqs.~\eqref{eq:EP_MLG_finite} and \eqref{eq:EP_BLG_finite}.  These are direct non-equilibrium rate-sum calculations.

The frequency correction is generally not symmetric in $\Delta T$ in the EP regime because the electronic phase-space factor is evaluated at $T_{\rm e}$, while the classical phonon amplitude is proportional to $T_{\rm ph}$.  In the BG zero-frequency limit the symmetric rate sum is even in $\Delta T$, as shown by Eq.~\eqref{eq:noneq_BG_expansion}, and at finite frequency the detailed curvature is determined by the shifted electron phase-space functions.

The strongest crossover test is obtained by evaluating the full doped-intraband MLG on-shell spectrum described in Sec.~\ref{sec:finite_results}.  This calculation keeps the exact triangle geometry, the MLG chirality factor, the Bose and Fermi distributions, and both frequency signs and phonon processes.  It does not impose $q\ll k_{\rm F}$, replace the Bose function by its classical form, or make the scattering quasi-elastic.

\begin{figure*}[t]
 \centering
 \includegraphics[width=\textwidth]{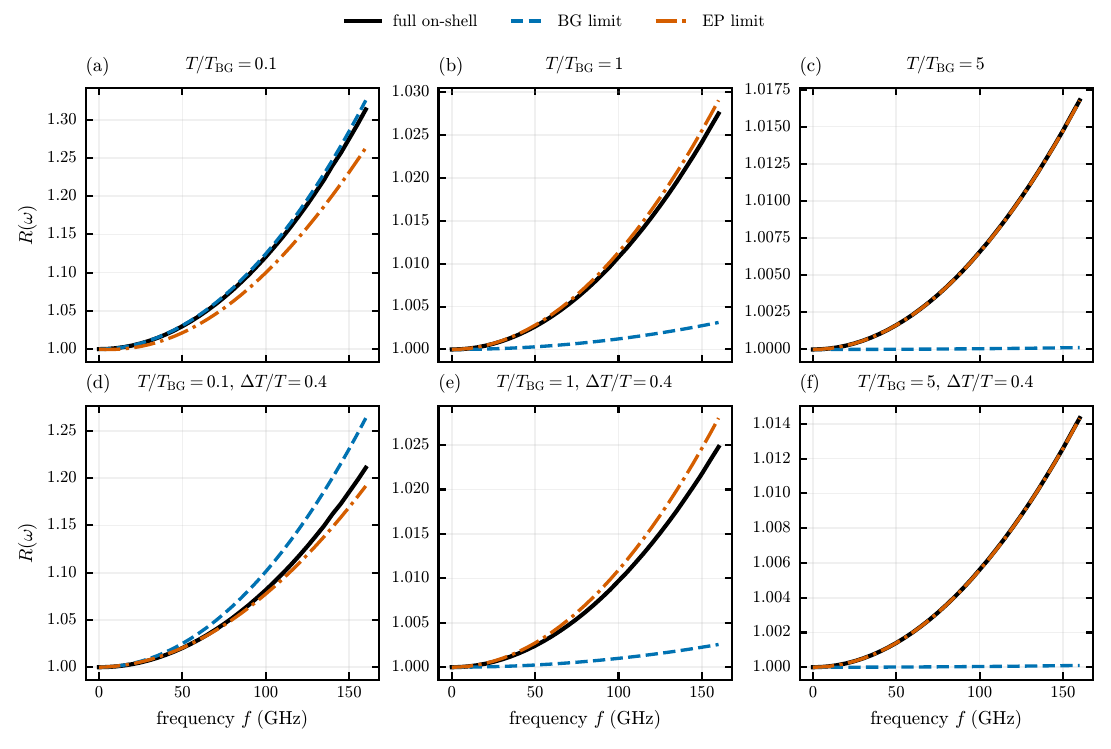}
 \caption{Full doped-intraband MLG finite-frequency ratio with the BG and EP reductions from
Eqs.~\eqref{eq:app_full_integral}
and \eqref{eq:R_definition}, compared with
the reduced BG and EP ratios obtained from
Eqs.~\eqref{eq:BG_sym_reduced}
and \eqref{eq:EP_MLG_finite}.  The upper row is equilibrium; the lower row uses $\Delta T/T=0.4$.  Columns show $T/T_{\rm BG}=0.1$, $1$, and $5$.  The full result tends to BG at low temperature and to EP at high temperature, while the crossover requires the unreduced on-shell integral.}
 \label{fig:full_finite}
\end{figure*}

Fig.~\ref{fig:full_finite} compares the normalized full result with the reduced BG and EP spectra at $T/T_{\rm BG}=0.1$, $1$, and $5$, for equilibrium and for $\Delta T/T=0.4$.  Deep in the BG regime the full curve follows the BG prediction.  At high temperature it follows the EP prediction.  Around $T_{\rm BG}$ neither limiting expression is uniformly accurate, and the full integral is required.  The same trend survives at finite temperature bias.  Thus, the BG and EP formulas are controlled asymptotic limits of one microscopic on-shell spectrum rather than unrelated models.

\begin{figure*}[t]
 \centering
 \includegraphics[width=0.96\textwidth]{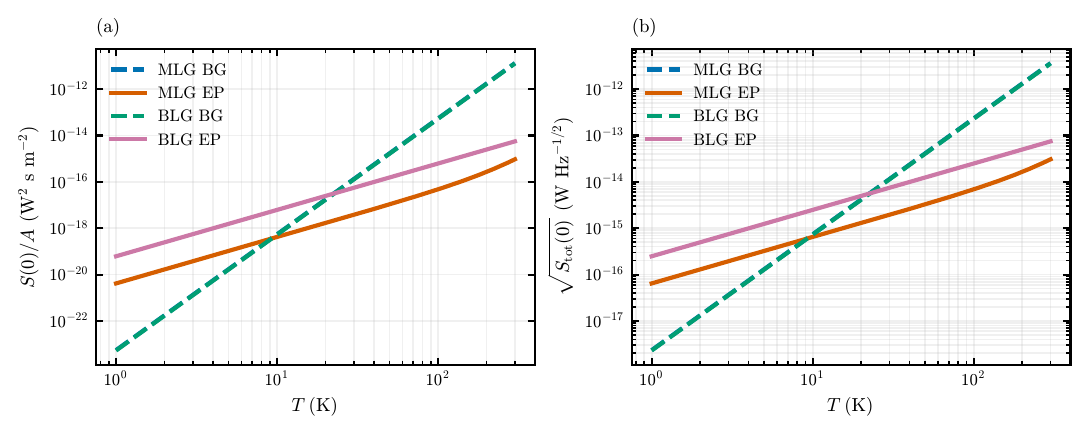}
 \caption{Absolute equilibrium zero-frequency noise. The BG curves follow
Eqs.~\eqref{eq:S0_MLG_BG}
and \eqref{eq:S0_BLG_BG}.
The EP curves are obtained from
Eqs.~\eqref{eq:G_MLG_EP}
and \eqref{eq:G_BLG_EP} using
$S(0)/A=2k_{\rm B}T^2[G_{\rm ep}(T)/A]$.  (a) Spectral density per area.  (b) Square-root total noise for a device area
$A=1\,\mu\mathrm{m}^2$.
The curves represent limiting BG and EP
expressions; extensions outside their
respective validity regimes are extrapolations.}
 \label{fig:absolute_scales}
\end{figure*}

The noise spectral density per area has dimensions \([S/A]=\mathrm{W^2\,s\,m^{-2}}\).
For a device of area $A_{\rm dev}$, the total spectrum is $S_{\rm tot}=A_{\rm dev}(S/A)$.  Fig.~\ref{fig:absolute_scales} gives both the per-area spectrum and the square-root device scale for $A_{\rm dev}=1\ \mu\mathrm{m}^2$.  The absolute values are small in SI units because individual phonon energies are microscopic and the BG noise is strongly suppressed as $T^5$.  Nevertheless, the corresponding amplitude can enter the femtowatt-per-root-hertz range in micron-scale devices, which is relevant to fast calorimetry and sensitive thermal detection~\cite{Fong2012,Gasparinetti2015,Saira2016}.

\section{Discussion}
\label{sec:discussion}


The average cooling power and the heat-current noise are two moments of the same microscopic energy-transfer process. The first moment reproduces the quasiequilibrium laws of Viljas and Heikkil\"a~\cite{Viljas2010}; the second moment contains additional information about fluctuations and associated time (energy) scales.  This relation is particularly transparent at zero frequency.  The current is proportional to the difference of emission and absorption rates, whereas the noise is proportional to their sum.  Detailed balance therefore sets the average current to zero at equilibrium without suppressing the exchange noise.



The EP limits are controlled mainly by electronic phase space.  The sound velocity cancels from the zero-frequency EP conductance.  MLG has a fourth-order electronic moment and therefore a strong neutral-regime temperature dependence.  BLG has a constant low-energy density of states and instead produces the weaker factor $T\ln[2\cosh(\mu/2k_{\rm B}T)]$.  These differences provide the possibility to distinguish mono- and bilayer cooling through their temperature and doping dependence.


The finite-frequency spectrum probes the correlation time of the energy exchange.  The detector energy $\hbar\omega$ enters both the energy-conservation shells and the current vertices.  The resulting factors $(\hbar\omega\mp2E_q)^2$ are absent from the static cooling-power problem.  They generate nonzero quantum noise even when the classical zero-frequency noise vanishes.  In the BG limit, Eq.~\eqref{eq:BG_zero_temperature} gives $S(\omega)\propto |\omega|^5$ at zero temperature.  This is consistent with the broader observation that finite-frequency heat noise contains quantum terms beyond the classical zero-frequency FDT~\cite{Averin2010,Pekola2018,IdrisovAnyonicHeat2024}. For this reason, the equilibrium identity $S(0)=2k_{\rm B}T^2G_{\rm ep}(T)$ should not be extended automatically to nonzero frequency.  In particular, a temperature difference is not a microscopic dynamical force in the same way as an electric field.  A carefully defined dynamic thermal response may be introduced for a specified protocol, but its relation to the full heat-current spectrum is not universal.  The present work therefore treats the finite-frequency spectrum as the primary object and derives it directly from the current operator.

The reduced BG and EP spectra give useful analytic understanding.  The full on-shell MLG calculation demonstrates where those approximations are valid and how they connect.  The numerical comparison also shows that the finite-frequency crossover depends on both $T/T_{\rm BG}$ and $hf/k_{\rm B}T$.  A system may be in the static EP regime while still requiring the full frequency dependence when $hf$ is no longer small compared with the electronic thermal scale.


When $T_{\rm e} \neq T_{\rm ph}$, the electron and phonon factors are governed by different temperatures.  In the BG regime the zero-frequency rate sum is $n_{\rm e}+n_{\rm ph}+2n_{\rm e} n_{\rm ph}$.  In the EP regime the classical phonon population supplies an overall $T_{\rm ph}$ factor while the electronic moments are evaluated at $T_{\rm e}$.  These structures explain why the noise is not determined by a single effective temperature.



A natural observable is the ratio $R(\omega)=S(\omega)/S(0)$, analogous to the normalized spectrum used in electron--phonon heat-noise studies of metals~\cite{Pekola2018}.  In the parameter range considered here, the correction is usually very small at a few gigahertz and becomes percent-level at tens to hundreds of gigahertz, with the largest relative correction occurring at low temperature.  The required frequency range is challenging but is moving toward the bandwidth of radio-frequency thermometry and circuit-based calorimetry~\cite{Gasparinetti2015,Saira2016}. The absolute scale depends strongly on $D^2$, device area, chemical potential, and the phonon environment.  An experimental comparison should therefore determine or independently constrain the deformation potential and account for the actual mass loading and acoustic mode structure.  The present results describe the intrinsic clean LA-phonon channel and provide a baseline against which additional cooling mechanisms can be identified.


It is worth mentioning, that the model omits several effects that can be important in specific samples.  Screening modifies the small-$q$ matrix element and may change BG power laws.  Disorder-assisted supercollisions can dominate the cooling over an intermediate temperature range~\cite{Song2012,Betz2013,Graham2013}.  Suspended devices can couple to flexural modes, while supported devices can couple to substrate and remote surface phonons.  Optical phonons dominate at sufficiently high temperature~\cite{Viljas2010}.  Electron--electron interactions are included only indirectly through the assumption of a local Fermi distribution. The full on-shell benchmark is presently restricted to the doped MLG intraband channel.  A complete calculation near charge neutrality would require both bands and a consistent ultraviolet cutoff for the low-energy model.  The analogous full BLG crossover should include the exact parabolic kinematics and, at sufficiently small momentum, possible interband processes.  These extensions are numerically more demanding and are left for future work.

\section{Conclusion}
\label{sec:conclusion}
We developed a microscopic theory of electron--phonon heat-current noise in monolayer and bilayer graphene. It is worth mentioning that the equilibrium FDT identity, $S(0)=2k_{\rm B}T^2G_{\rm ep}(T)$, can not be applied straightforwardly for finite frequencies as in case of electric current. To be more precise, the temperature difference is not a microscopic dynamical force in the same way as an electric field. Thus, the primary focus of present work is the finite-frequency spectrum which we calculated using the direct of heat current operator between electron and phonon subsystems.

Starting from the deformation-potential Hamiltonian, we derived a symmetric subsystem heat current operator and evaluated its leading current--current correlator.  The direct Wick contraction vanishes because momentum conservation forces an acoustic phonon with zero momentum and zero energy.  The exchange contraction produces the physical emission and absorption spectrum.

At zero frequency, both processes carry the same transferred-energy weight and must be retained.  The resulting equilibrium spectra satisfy the thermodynamic relation $S(0)=2k_{\rm B}T^2G_{\rm ep}(T)$.  This reproduces the Viljas--Heikkil\"a conductance laws from Ref.~\cite{Viljas2010} and yields the noise scalings $S^{\rm MLG,BG}(0)\propto |\mu|T^5$ and $S^{\rm BLG,BG}(0)\propto |\mu|^{-1/2}T^5$.  The EP results reflect the different electronic phase spaces of MLG and BLG.

At finite frequency, the current vertices become $(\hbar\omega \mp 2E_q)^2$. We derived reduced BG and EP spectra for separate electron and phonon temperatures and evaluated them without a low-frequency expansion.  The corresponding quadratic asymptotics agree at small frequency with numerics. A full doped-intraband MLG on-shell calculation connects the two regimes and shows explicitly that the BG and EP spectra are the low- and high-temperature limits of one microscopic result.

Out of equilibrium, the noise is a rate sum governed separately by $T_{\rm e}$ and $T_{\rm ph}$.  It is not fixed by FDT and should not be interpreted as a unique nonlinear thermal conductance.  The finite-frequency ratios and heat maps proposed here provide direct signatures of this non-equilibrium energy-exchange statistics.

The present theory supplies an intrinsic acoustic-phonon baseline for graphene calorimetry and thermal-noise spectroscopy.

\begin{acknowledgments}
This work is supported by United Arab Emirates University under the Startup
Grant No. G00004974, UPAR Grant G00005479 and UAEU-AUA Grant G00005522.
\end{acknowledgments}



\appendix
\section{Heat-current operator and Wick expansion}
\label{app:wick}

This appendix gives the algebra behind Eqs.~\eqref{eq:heat_current_operator} and \eqref{eq:general_unsym_kernel}.

\subsection{Subsystem energy derivatives}
Only $\hat{H}_{\rm e\text{-}{\rm ph}}$ contributes to the subsystem energy derivatives.  The fermionic commutator is \( [\hat{c}_{{\bf p}\beta}^{\dagger}\hat{c}_{{\bf k}\alpha},\hat{H}_{\rm e}]
 =\left(\epsilon_{{\bf k}\alpha}-\epsilon_{{\bf p}\beta}\right)
 \hat{c}_{{\bf p}\beta}^{\dagger}\hat{c}_{{\bf k}\alpha}\) and the bosonic commutators are \([\hat{b}_{\bf q},\hat{H}_{\rm ph}]=E_q \hat{b}_{\bf q}\), \([\hat{b}_{-{\bf q}}^{\dagger},\hat{H}_{\rm ph}]=-E_q \hat{b}_{-{\bf q}}^{\dagger}\). Substitution into Eq.~\eqref{eq:current_definition} gives Eq.~\eqref{eq:heat_current_operator} from main text.

\subsection{Free evolution and contractions}

To second order in the electron-phonon matrix element, all operators may be evolved with $\hat{H}_0=\hat{H}_{\rm e}+\hat{H}_{\rm ph}$,
\begin{align}
 \hat{c}_{{\bf k}\alpha}(t)&=\hat{c}_{{\bf k}\alpha}e^{-i\epsilon_{{\bf k}\alpha}t/\hbar},
 &\hat{c}_{{\bf k}\alpha}^{\dagger}(t)&=\hat{c}_{{\bf k}\alpha}^{\dagger}e^{i\epsilon_{{\bf k}\alpha}t/\hbar},\\
 \hat{b}_{\bf q}(t)&=\hat{b}_{\bf q}e^{-iE_qt/\hbar},
 &\hat{b}_{\bf q}^{\dagger}(t)&=\hat{b}_{\bf q}^{\dagger}e^{iE_qt/\hbar}.
\end{align}
The nonzero phonon contractions are
\begin{align}
 \langle \hat{b}_{\bf q}(t)\hat{b}_{-{\bf q}'}^{\dagger}(0)\rangle
 &=[1+n_q]e^{-iE_qt/\hbar}\delta_{{\bf q},-{\bf q}'},\\
 \langle \hat{b}_{-{\bf q}}^{\dagger}(t)\hat{b}_{{\bf q}'}(0)\rangle
 &=n_qe^{iE_qt/\hbar}\delta_{{\bf q},-{\bf q}'}.
\end{align}
The electron four-point correlation function is
\begin{align}
 &\langle \hat{c}_{{\bf p}\beta}^{\dagger}\hat{c}_{{\bf k}\alpha}
 \hat{c}_{{\bf p}'\beta'}^{\dagger}\hat{c}_{{\bf k}'\alpha'}\rangle
 \nonumber\\
 &\quad=f_{{\bf p}\beta}f_{{\bf p}'\beta'}
 \delta_{{\bf p}{\bf k}}\delta_{\beta\alpha}
 \delta_{{\bf p}'{\bf k}'}\delta_{\beta'\alpha'}
 \nonumber\\
 &\qquad+f_{{\bf p}\beta}[1-f_{{\bf k}\alpha}]
 \delta_{{\bf p}{\bf k}'}\delta_{\beta\alpha'}
 \delta_{{\bf k}{\bf p}'}\delta_{\alpha\beta'}.
 \label{eq:app_wick}
\end{align}
The first term is the direct contraction.  It imposes ${\bf p}={\bf k}$, and translational invariance then imposes ${\bf q}=0$.  Because $E_{q=0}=0$ and $\Delta_{kk}=0$, both current vertices vanish.  Thus the direct term is zero before any frequency integration.

The exchange contraction in Eq.~\eqref{eq:app_wick}
pairs the creation operator in each electronic
bilinear with the annihilation operator in the
other. Its Kronecker deltas impose
${\bf k}'={\bf p}$, ${\bf p}'={\bf k}$,
$\alpha'=\beta$, and $\beta'=\alpha$.
Thus, the two current vertices connect the same
pair of electronic states in opposite directions.
Momentum conservation at the vertices then gives
${\bf q}'=-{\bf q}$.

Hermiticity of the electron--phonon interaction
implies
\(M_{{\bf k}{\bf p},-{\bf q}}^{\alpha\beta}
=\bigl(M_{{\bf p}{\bf k},{\bf q}}^{\beta\alpha}\bigr)^{\ast}\), which means that the matrix element for the reverse transition
is the complex conjugate of that for the
forward transition.
Consequently, the product of matrix elements
from the two current vertices becomes
\begin{equation}
 M_{{\bf p}{\bf k},{\bf q}}^{\beta\alpha}
 M_{{\bf k}{\bf p},-{\bf q}}^{\alpha\beta}
 =
 \left|M_{{\bf p}{\bf k},{\bf q}}^{\beta\alpha}\right|^2.
\end{equation}
The exchange contribution therefore weights each
allowed electronic transition by its squared
coupling amplitude and the occupation factor
$f_{{\bf p}\beta}[1-f_{{\bf k}\alpha}]$.
This is the transition weight entering the
emission and absorption terms of
Eq.~\eqref{eq:general_unsym_kernel}. Using the Fourier identity \(\int dt
 \exp[i(\hbar\omega-X)t/\hbar]
 =2\pi\hbar\,\delta(\hbar\omega-X)\) with \(X=\Delta_{kp}^{\alpha\beta}\pm E_q\) and combining the electronic evolution with the two
phonon contractions gives the time-dependent factors
$\exp[-i(\Delta_{kp}^{\alpha\beta}+E_q)t/\hbar]$
and
$\exp[-i(\Delta_{kp}^{\alpha\beta}-E_q)t/\hbar]$ and therefore results 
$\delta(\hbar\omega-\Delta_{kp}^{\alpha\beta}-E_q)$
and
$\delta(\hbar\omega-\Delta_{kp}^{\alpha\beta}+E_q)$,
respectively, yielding the two energy-conservation delta functions in
Eq.~\eqref{eq:general_unsym_kernel} from main text.

\subsection{Zero-frequency shell weights}

The two energy-conservation delta functions derived
above determine the current-vertex weights in the
zero-frequency limit. Setting $\omega=0$ in
Eq.~\eqref{eq:general_unsym_kernel} gives
\begin{align}
 \delta(-\Delta-E_q)
 &\Rightarrow \Delta=-E_q,
 &(\Delta-E_q)^2&=4E_q^2,\\
 \delta(-\Delta+E_q)
 &\Rightarrow \Delta=E_q,
 &(\Delta+E_q)^2&=4E_q^2,
\end{align}
where $\Delta\equiv\Delta_{kp}^{\alpha\beta}$.
Thus, both phonon emission and absorption acquire
the same squared current-vertex weight, $4E_q^2$,
although their occupation factors generally differ.
Multiplying this weight by the prefactor
$\pi/(2\hbar)$ in
Eq.~\eqref{eq:general_unsym_kernel} gives
$2\pi/\hbar$. Since the symmetrized and
unsymmetrized spectra coincide at $\omega=0$,
the resulting expression is the emission--absorption
rate sum in Eq.~\eqref{eq:zero_kernel}.
At equilibrium and finite temperature, both rates
contribute to the noise; omitting absorption would
give an incomplete rate sum and violate the
zero-frequency FDT.
\section{Bloch--Gr\"uneisen derivations}
\label{app:bg}

\subsection{Monolayer graphene}
For a doped conduction band, let $\theta$ denote
the angle between the electron wave vector ${\bf k}$
and the phonon wave vector ${\bf q}$.
In the BG regime $q\ll k_{\rm F}$, thus
\begin{equation}
 \epsilon_{\bf k}-\epsilon_{{\bf k}-{\bf q}}
 \simeq \hbar v_{\rm F} q\cos\theta,
 \qquad F_{++}^{\rm MLG}\simeq1.
\end{equation}
The angular delta function is
\begin{align}
 \int_0^{2\pi}d\theta\,
 \delta[\hbar q(v_{\rm F}\cos\theta-v_{\rm s})]
 &=\frac{2}{\hbar v_{\rm F} q\sqrt{1-(v_{\rm s}/v_{\rm F})^2}}\nonumber\\
 &\simeq\frac{2}{\hbar v_{\rm F} q}.
 \label{eq:app_mlg_bg_angle}
\end{align}
After converting the momentum sums to integrals, Eq.~\eqref{eq:zero_kernel} becomes
\begin{equation}
 S^{\rm MLG}_{\rm BG}(0)=
 \frac{g_dAD^2\hbar v_{\rm s}}{4\pi^2\rho_{\rm MLG}v_{\rm F}}
\int_0^{\infty}dk\,k\int_0^{\infty}dq\,q^3{\cal F}(k,q),
 \label{eq:app_mlg_reduced}
\end{equation}
where ${\cal F}$ contains the two Fermi--Bose rate factors
\begin{align}
 {\cal F}(k,q)
 &=f_{\rm e}(\epsilon_k)
 [1-f_{\rm e}(\epsilon_k-E_q)](1+n_q)\nonumber\\
 &\quad+f_{\rm e}(\epsilon_k-E_q)
 [1-f_{\rm e}(\epsilon_k)]n_q,
\end{align}
with $E_q=\hbar v_{\rm s} q$,
$f_{\rm e}(\epsilon)=[\exp[(\epsilon-\mu)/(k_{\rm B}T_{\rm e})]+1]^{-1}$,
and $n_q=[\exp[E_q/(k_{\rm B}T_{\rm ph})]-1]^{-1}$.
The first term describes phonon emission and the
second describes the reverse absorption process.

Near the Fermi surface, $kdk \simeq (k_{\rm F}/\hbar v_{\rm F})d\epsilon$. Thus, keeping the finite phonon energy $E_q$ inside the Fermi functions gives
\begin{align}
 \int d\epsilon\,f_{\rm e}(\epsilon+E_q)[1-f_{\rm e}(\epsilon)]&=E_qn_{\rm B}(E_q),\\
 \int d\epsilon\,f_{\rm e}(\epsilon)[1-f_{\rm e}(\epsilon+E_q)]&=E_q[1+n_{\rm B}(E_q)].
\end{align}
Hence $\int d\epsilon\,{\cal F}=2E_qn_q(1+n_q)$ and using $E_q=\hbar v_{\rm s}q$ in Eq.~\eqref{eq:app_mlg_reduced} gives
\begin{equation}
 S^{\rm MLG}_{\rm BG}(0)=
 \frac{g_dAD^2k_{\rm F}\hbar v_{\rm s}^2}{2\pi^2\rho_{\rm MLG}v_{\rm F}^2}
 \int_0^{\infty}dq\,q^4n_q(1+n_q).
\end{equation}
After performing the integration by parts we have
\begin{equation}
 \int_0^{\infty}dq\,q^4n_q(1+n_q)
 =\frac{4\pi^4}{15}\left(\frac{k_{\rm B}T}{\hbar v_{\rm s}}\right)^5.
\end{equation}
Thus, the final result is 
\begin{equation}
 S^{\rm MLG}_{\rm BG}(0)=
 \frac{2g_d\pi^2}{15}
 \frac{AD^2|\mu|k_{\rm B}^5T^5}
 {\rho_{\rm MLG}\hbar^5v_{\rm F}^3v_{\rm s}^3},
\end{equation}
which reduces to Eq.~\eqref{eq:S0_MLG_BG} for $g_d=4$.

\subsection{Bilayer graphene}
For BLG, the spectrum and Fermi velocity $\epsilon_k=\hbar^2k^2/2m$ and $u_{\rm F}=\hbar k_{\rm F}/m$, thus at small $q$,
\begin{equation}
 \epsilon_{\bf k}-\epsilon_{{\bf k}-{\bf q}}
 =\frac{\hbar^2}{2m}(2kq\cos\theta-q^2)
 \simeq\hbar u_{\rm F}q\cos\theta,
\end{equation}
and $F_{++}^{\rm BLG}=\cos^2\phi\simeq1$.  The angular factor is therefore $2/(\hbar u_{\rm F}q)$ and the radial measure is $kdk=(m/\hbar^2)d\epsilon$. Repeating the MLG steps gives
\begin{equation}
 S^{\rm BLG}_{\rm BG}(0)=
 \frac{2g_d\pi^2}{15}
 \frac{AD^2m}{\rho_{\rm BLG}u_{\rm F}}
 \frac{k_{\rm B}^5T^5}{\hbar^5v_{\rm s}^3}.
\end{equation}
Since $m/u_{\rm F}=(\gamma_1/4v^3_0)\sqrt{\gamma_1/|\mu|}$, we obtain Eq.~\eqref{eq:S0_BLG_BG} for $g_d=4$.

\subsection{Nonequilibrium Bose integral}

For $T_{\rm e}\neq T_{\rm ph}$ we evaluate the electronic
energy integrals in the emission--absorption rate sum
of Eq.~\eqref{eq:zero_kernel}, using the BG
Fermi-surface reduction described above.
The emission integral gives $En_{\rm e}(E)$, whereas the
absorption integral gives $E[1+n_{\rm e}(E)]$.
Including the corresponding phonon occupation
factors therefore yields \(E\left[n_{\rm e}(1+n_{\rm ph})+(1+n_{\rm e})n_{\rm ph}\right]
 =E[n_{\rm e}+n_{\rm ph}+2n_{\rm e}n_{\rm ph}]\)
and using the symmetric temperatures
\(T_{\rm e}=T(1+r)\) and \(T_{\rm ph}=T(1-r)\), with
\(r=\Delta T/(2T)\), and introducing \(x=E/k_{\rm B}T\),
we define the dimensionless phonon-energy integral
\begin{align}
\mathcal{L}_{\rm BG}(r)&=\int_0^{\infty}dx\,x^4
 \left[n\!\left(\frac{x}{1+r}\right)
 +n\!\left(\frac{x}{1-r}\right)\right.\nonumber\\
 &\left.\quad+2n\!\left(\frac{x}{1+r}\right)
 n\!\left(\frac{x}{1-r}\right)\right],
 \label{eq:app_JBG}
\end{align}
where $n(y)=(e^y-1)^{-1}$.
Expanding with the Bose series
$n(y)=\sum_{l\ge1}e^{-ly}$ gives
\begin{equation}
\mathcal{L}_{\rm BG}(r)=\frac{8\pi^4}{15}
 +\frac{32\pi^6}{63}r^2+O(r^4).
\end{equation}

Within the BG approximation, the material-dependent
prefactor cancels when the nonequilibrium noise is
normalized to its equilibrium value at the same mean
temperature $T$ and fixed chemical potential:
\begin{equation}
 \frac{S_{\rm BG}(0;T_e,T_{\rm ph})}
      {S_{\rm BG}(0;T,T)}
 =\frac{\mathcal{L}_{\rm BG}(r)}{\mathcal{L}_{\rm BG}(0)}.
\end{equation}
Using $\mathcal{L}_{\rm BG}(0)=8\pi^4/15$ and
$r=\Delta T/(2T)$ recovers
Eq.~\eqref{eq:noneq_BG_expansion} in the main text.
The absence of a linear term follows from the
invariance of the BG rate sum under
$T_{\rm e}\leftrightarrow T_{\rm ph}$.

\section{Equipartition derivations}
\label{app:ep}

\subsection{Common quasi-elastic reduction}
Starting from the zero-frequency rate sum in
Eq.~\eqref{eq:zero_kernel}, we consider equilibrium,
$T_{\rm e}=T_{\rm ph}=T$, in the EP regime.
The phonon occupations satisfy
$n_q\simeq k_{\rm B}T/E_q\gg1$ and
$1+n_q\simeq n_q$.
In the quasi-elastic approximation, we neglect $E_q$
in both electronic energy-conservation delta functions,
while retaining it in the heat-current weights:
\begin{equation}
 \delta(\epsilon_{k\alpha}-\epsilon_{p\alpha}\pm E_q)
 \simeq\delta(\epsilon_{k\alpha}-\epsilon_{p\alpha}).
\end{equation}
Both electronic occupation products therefore reduce
to $f_{\rm e}(1-f_{\rm e})$ and
\begin{align}
 E_q^2f_{\rm e}(1-f_{\rm e})(1+2n_q)
 &\simeq2k_{\rm B}TE_qf_{\rm e}(1-f_{\rm e})\nonumber\\
 &=2k_{\rm B}^2T^2E_q
 \left(-\frac{\partial f_{\rm e}}{\partial\epsilon}\right).
\end{align}
This produces the microscopic FDT pair
\begin{align}
 S_{\rm EP}(0)
 &=\frac{4\pi k_{\rm B}^2T^2}{\hbar}g_d
 \sum_{kp\alpha}|M_{pk}|^2E_q\nonumber\\
 &\quad\times\left(-\frac{\partial f_{\rm e}}{\partial\epsilon}\right)
 \delta(\epsilon_{k\alpha}-\epsilon_{p\alpha}),\\
 G_{\rm EP}(T)
 &=\frac{2\pi k_{\rm B}}{\hbar}g_d
 \sum_{kp\alpha}|M_{pk}|^2E_q\nonumber\\
 &\quad\times\left(-\frac{\partial f_{\rm e}}{\partial\epsilon}\right)
 \delta(\epsilon_{k\alpha}-\epsilon_{p\alpha}).
\end{align}
Their ratio is $2k_{\rm B}T^2$ before the remaining integrals are evaluated.

\subsection{Monolayer graphene}

We now evaluate the remaining momentum integral in
Eq.(C4) for MLG
to obtain Eq.~\eqref{eq:G_MLG_EP}, namely 
\begin{equation}
 |M_{pk}|^2E_q=\frac{D^2\hbar^2q^2}{2A\rho_{\rm MLG}}F(\phi),
\end{equation}
so $v_{\rm s}$ cancels.  Elastic intraband scattering gives
\begin{equation}
 \delta(\epsilon_{k\alpha}-\epsilon_{p\alpha})
 =\frac{1}{\hbar v_{\rm F}}\delta(k-p).
\end{equation}
On the shell $p=k$,
\begin{equation}
 q^2F_{++}^{\rm MLG}=2k^2(1-\cos\phi)\frac{1+\cos\phi}{2}
 =k^2\sin^2\phi.
\end{equation}
The angular part of the momentum integral in
Eq.(C4) is therefore
\begin{equation}
 \int_0^{2\pi}d\theta_k\int_0^{2\pi}d\theta_p\,
 q^2F_{++}^{\rm MLG}
 =2\pi^2k^2,
\end{equation}
where $\phi=\theta_k-\theta_p$, and thus 
\begin{equation}
 \frac{G_{\rm EP}^{\rm MLG}}{A}
 =\frac{g_dD^2k_{\rm B}}{8\pi\rho_{\rm MLG}\hbar^5v_{\rm F}^6}
 \int d\epsilon\,\epsilon^4
 \left(-\frac{\partial f_{\rm e}}{\partial\epsilon}\right).
\end{equation}
The exact Fermi-window moment is
\begin{equation}
 \int d\epsilon\,\epsilon^4(-f'_{\rm e})
 =\mu^4+2\pi^2\mu^2(k_{\rm B}T)^2
 +\frac{7\pi^4}{15}(k_{\rm B}T)^4,
\end{equation}
leading to Eq.~\eqref{eq:G_MLG_EP}.

\subsection{Bilayer graphene}
To obtain the BLG equipartition conductance
in Eq.~\eqref{eq:G_BLG_EP}, we start from the
general conductance expression, Eq.~(C4).
We evaluate its momentum integrals using the
parabolic BLG dispersion and the intraband
chirality factor $F=\cos^2\phi$, where $\phi$
is the angle between ${\bf k}$ and ${\bf p}$.
Elastic energy conservation fixes $p=k$,
so the radial delta function becomes
\begin{equation}
 \delta(\epsilon_k-\epsilon_p)=\frac{m}{\hbar^2p}\delta(p-k).
\end{equation}
The intraband chirality factor is $F=\cos^2\phi$.  On $p=k$,
\begin{equation}
 q^2F=2k^2(1-\cos\phi)\cos^2\phi,
\end{equation}
and
\begin{equation}
 \int d\theta_kd\theta_p\,q^2F=4\pi^2k^2.
\end{equation}
The conductance then becomes
\begin{equation}
 \frac{G_{\rm EP}^{\rm BLG}}{A}
 =\frac{g_dD^2k_{\rm B}m^3}{2\pi\rho_{\rm BLG}\hbar^5}
 \int dE\,|E|\left(-\frac{\partial f_{\rm e}}{\partial E}\right).
\end{equation}
Splitting the signed-energy integral into electron and hole parts gives
\begin{equation}
 \int dE\,|E|(-f'_{\rm e})=2k_{\rm B}T\ln\left[2\cosh\left(\frac{\mu}{2k_{\rm B}T}\right)\right].
\end{equation}
Using the above two equations and $m=\gamma_1/(2v_0^2)$ and $g_d=4$ gives Eq.~\eqref{eq:G_BLG_EP} from main text.

\subsection{Two-temperature EP rate sum}

At different temperatures, the classical phonon factor is evaluated at $T_{\rm ph}$ while $f_{\rm e}(1-f_{\rm e})=k_{\rm B}T_{\rm e}(-\partial_\epsilon f_{\rm e})$. Consequently, the same momentum integral that defines $g(\mu,T_{\rm e})$ is multiplied by $2k_{\rm B}T_{\rm e}T_{\rm ph}$, yielding Eq.~\eqref{eq:noneq_EP_common}.  This direct calculation is the origin of that formula; it is not obtained by applying FDT away from equilibrium.

\section{Finite-frequency reductions and asymptotics}
\label{app:finite}

\subsection{BG continuum reduction}
In the doped BG regime, the electronic energy difference
$\Delta\equiv\epsilon_{{\bf k}\alpha}
-\epsilon_{{\bf p}\alpha}$, with ${\bf p}={\bf k}-{\bf q}$,
is linearized as $\Delta\simeq\hbar v_{\rm F}q\cos\theta$
for MLG. For the two finite-frequency shells, define
$c_\mp=(\hbar\omega\mp E_q)/(\hbar v_{\rm F}q)$.
The angular integral is
\begin{equation}
 \int_0^{2\pi}d\theta\,
 \delta(\hbar\omega-\hbar v_{\rm F} q\cos\theta\mp E_q)
 =
 \frac{2\Theta(1-|c_\mp|)}
 {\hbar v_{\rm F} q\sqrt{1-c_\mp^2}}.
\end{equation}
The leading BG continuum reduction replaces this
factor by $2/(\hbar v_{\rm F}q)$ when $|c_\mp|\ll1$.
Thus, the shell sign affects the exact Jacobian and
kinematic constraint, but drops out of its leading
approximation.
The Fermi-surface and phonon-energy measures are
$kdk \simeq (k_{\rm F}/\hbar v_{\rm F})d\epsilon$ and
$qdq =(E/\hbar^2 v_s^2)dE$.
The first frequency shell imposes $\Delta=\hbar\omega-E$
and gives the vertex $(\hbar\omega-2E)^2$ together with
\begin{equation}
 \int d\epsilon\,
 f_{\rm e}(\epsilon+E-\hbar\omega)
 [1-f_{\rm e}(\epsilon)]
 =\Phi_{\rm e}(E-\hbar\omega; T_{\rm e})
 \label{eq:BG_unsym_reduced}
\end{equation}
The second shell imposes $\Delta=\hbar\omega+E$ and gives
$(\hbar\omega+2E)^2\Psi_{\rm e}(E+\hbar\omega;T_{\rm e})$.
Substituting these shell weights and electronic
energy integrals into
Eq.~\eqref{eq:general_unsym_kernel} gives
Eq.~\eqref{eq:BG_unsym_reduced}.
Symmetrizing in $\omega$ then gives
Eq.~\eqref{eq:BG_sym_reduced}.
BLG has the same dimensionless kernel; its parabolic
radial measure and angular Jacobian change only the prefactor.

For the symmetric temperature bias,
$T_{\rm e}=T(1+r)$ and $T_{\rm ph}=T(1-r)$, where
$r=\Delta T/(2T)$, introduce
$x=E/(k_{\rm B}T)$ and
$\Omega=\hbar\omega/(k_{\rm B}T)$.
The dimensionless occupation and electronic
phase-space functions are
\begin{align}
 &N_{\rm ph}(x;r)=\frac{1}{e^{x/(1-r)}-1},\\
 &\Phi_r(z)=\frac{z}{e^{z/(1+r)}-1},\\
 &\Psi_r(z)=\frac{z}{1-e^{-z/(1+r)}}.
\end{align}
Under this substitution,
$\Phi_{\rm e}(E-\hbar\omega;T_{\rm e})
=k_{\rm B}T\,\Phi_r(x-\Omega)$ and
$\Psi_{\rm e}(E+\hbar\omega;T_{\rm e})
=k_{\rm B}T\,\Psi_r(x+\Omega)$.
Equation~\eqref{eq:BG_unsym_reduced} therefore
contains the dimensionless kernel
\begin{align}
 {\cal B}_r(x,\Omega)
 &=(\Omega-2x)^2[1+N_{\rm ph}(x;r)]\Phi_r(x-\Omega)
 \nonumber\\
 &\quad+(\Omega+2x)^2N_{\rm ph}(x;r)\Psi_r(x+\Omega).
 \label{eq:app_Br}
\end{align}
Combining the dimensionless kernel
${\cal B}_r(x,\Omega)$ in Eq.~\eqref{eq:app_Br}
with the measure $x\,dx$ and symmetrizing through
$[{\cal B}_r(x,\Omega)+{\cal B}_r(x,-\Omega)]/2$
gives Eq.~\eqref{eq:BG_dimensionless}.
The overall $(k_{\rm B}T)^5$ factor follows from
the energy measure $E\,dE$, the squared current
vertex, and the electronic phase-space function,
which contribute two, two, and one powers of
$k_{\rm B}T$, respectively.

\subsection{BG low-frequency coefficient}
\begin{figure*}[t]
 \centering
 \includegraphics[width=\textwidth]{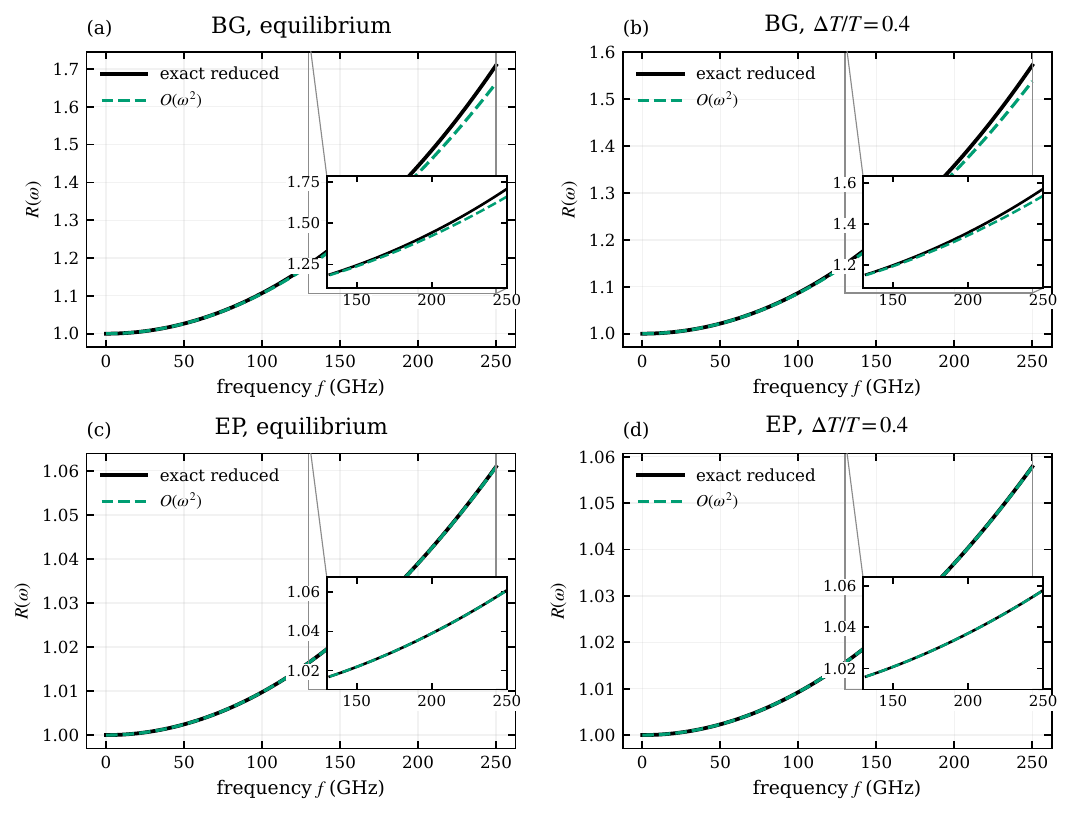}
\caption{Exact reduced MLG finite-frequency ratios
compared with their quadratic low-frequency
approximations.
The upper panels show BG results at mean
temperature $T=5$ K, calculated from
Eq.~\eqref{eq:BG_dimensionless}.
The equilibrium BG approximation follows
Eq.~\eqref{eq:BG_quadratic}; the nonequilibrium
quadratic coefficient is obtained from the
small-frequency expansion of the same BG integral,
as described in Appendix~\ref{app:finite}.
The lower panels show EP results at $T=100$ K
from Eq.~\eqref{eq:EP_MLG_finite},
with quadratic approximations from
Eq.~\eqref{eq:EP_MLG_quadratic}.
The left column is at equilibrium; the right
column uses $T_{\rm e/ph}=T\pm\Delta T/2$ with
$\Delta T/T=0.4$.
All ratios use Eq.~\eqref{eq:R_definition}.
Insets magnify the high-frequency separation.}
 \label{fig:appendix_exact_mlg}
\end{figure*}

\begin{figure*}[t]
 \centering
 \includegraphics[width=\textwidth]{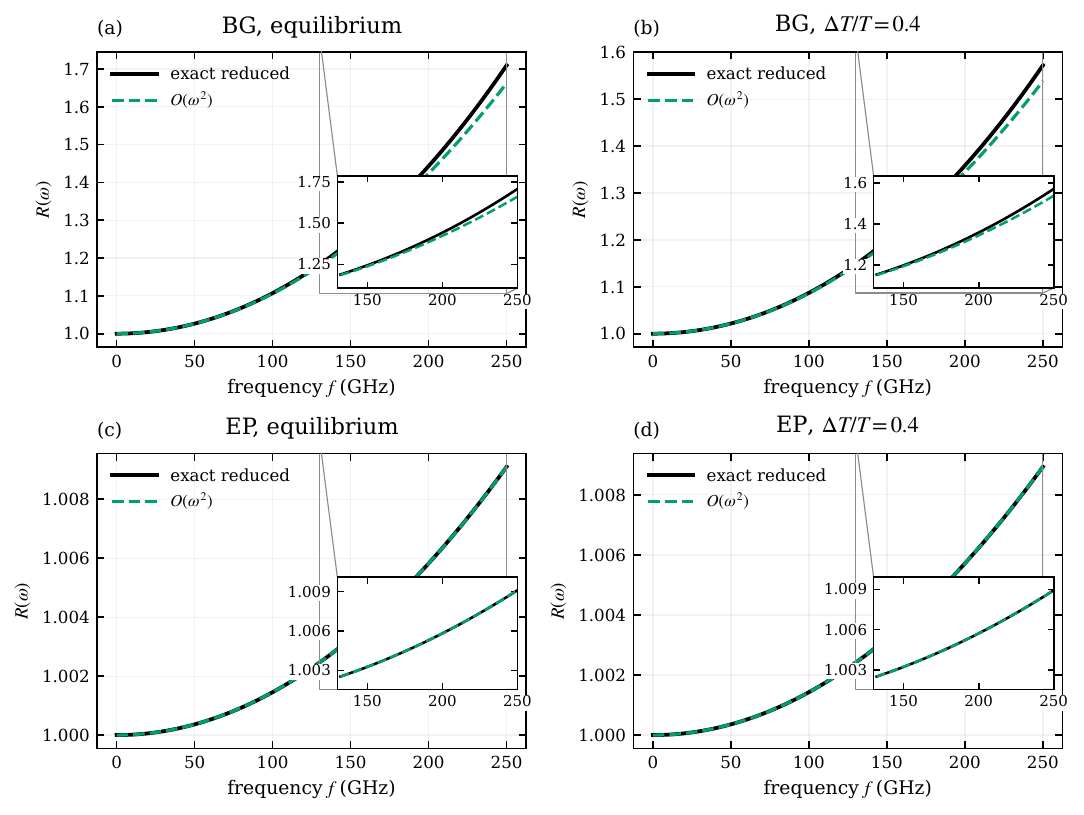}
 \caption{Exact reduced BLG finite-frequency ratios and quadratic approximations. The upper panels show BG results at mean
temperature $T=5$ K from
Eq.~\eqref{eq:BG_dimensionless}.
The equilibrium quadratic approximation follows
Eq.~\eqref{eq:BG_quadratic}; the nonequilibrium
coefficient is obtained from the small-frequency
expansion described in Appendix~\ref{app:finite}.  The BG ratio has the same dimensionless form as in MLG.  The BLG EP panels use the \(\mathcal{J}_0 \) and \(\mathcal{J}_1 \) energy integrals.}
 \label{fig:appendix_exact_blg}
\end{figure*}

To obtain the quadratic low-frequency correction,
we expand the dimensional BG kernel entering
Eq.~\eqref{eq:BG_unsym_reduced} about frequency,
at fixed $T_{\rm e}$ and $T_{\rm ph}$.
For compactness, write
$\Phi(E)=\Phi_{\rm e}(E;T_{\rm e})$,
$\Psi(E)=\Psi_{\rm e}(E;T_{\rm e})$, and
$N(E)=n_{\rm ph}(E,T_{\rm ph})$.
The kernel is \(
 {\cal B}(E,\hbar\omega)=(\hbar\omega-2E)^2[1+N(E)]
 \Phi(E-\hbar\omega)+(\hbar\omega+2E)^2N(E)
 \Psi(E+\hbar\omega)\).
Symmetrization cancels the odd powers of $\omega$,
giving \({\cal B}(E,\hbar \omega)+{\cal B}(E,-\hbar \omega)
 =2{\cal B}(E,0)+(\hbar \omega)^2{\cal B}_2(E)+O((\hbar \omega)^4)\),
where \(
 {\cal B}_2(E)=(1+N)\left[2\Phi+8E\Phi'+4E^2\Phi''\right]+N\left[2\Psi+8E\Psi'+4E^2\Psi''\right]\).
At equilibrium, Bose-series integration or repeated integration by parts gives
\begin{equation}
 \frac{1}{2}\frac{\int_0^{\infty}dE\,E{\cal B}_2(E)}
 {\int_0^{\infty}dE\,E{\cal B}(E,0)}
 =\frac{1}{(k_{\rm B}T)^2}
 \left(\frac{1}{12}+\frac{5}{16\pi^2}\right),
\end{equation}
which yields Eq.~\eqref{eq:BG_quadratic}.  Fig.~\ref{fig:appendix_exact_mlg} compares the complete reduced integrals with the quadratic approximations.  The inset magnifies the range in which the curves begin to separate.
The same comparison for BLG is shown in Fig.~\ref{fig:appendix_exact_blg}.  The BG normalized kernel is common to MLG and BLG, while the EP electronic phase space is different.

\subsection{Zero-temperature BG spectrum}

At $T_{\rm e}=T_{\rm ph}=0$ and $\hbar \omega>0$, the only nonvanishing interval in the symmetrized BG integral is $0<E<\hbar \omega$.  Direct evaluation gives
\begin{equation}
 \frac{1}{2}\int_0^{\hbar \omega} dE\,E(\hbar\omega-E)(\hbar \omega-2E)^2=\frac{(\hbar \omega)^5}{60},
\end{equation}
which yields Eq.~\eqref{eq:BG_zero_temperature}.  The $|\omega|^5$ law is the graphene BG counterpart of the finite-frequency quantum heat noise found in metallic systems~\cite{Pekola2018}.

\subsection{MLG EP angular factors}

In the quasi-elastic EP reduction, the finite electron energies are $\epsilon$ and $\epsilon+\hbar \omega$.  The magnitudes of the two electron momenta are $|\epsilon|/(\hbar v_{\rm F})$ and $|\epsilon+\hbar \omega|/(\hbar v_{\rm F})$.  The angular averages are
\begin{align}
 \int d\theta_kd\theta_p\,F^{\rm MLG}(\phi)&=2\pi^2,\\
 \int d\theta_kd\theta_p\,q^2F^{\rm MLG}(\phi)
 &=2\pi^2\left(k^2+p^2-kp\right).
\end{align}
They produce the two energy integrals \(\mathcal{I}_2\) and \(\mathcal{I}_4\) in Eqs.~\eqref{eq:I2andI4}.  At $\hbar \omega=0$,
\(
 \mathcal{I}_4(0;T_e)=2k_{\rm B}T_{\rm e} M_4(\mu,T_e)\),
which reduces Eq.~\eqref{eq:EP_MLG_finite} to Eq.~\eqref{eq:noneq_EP_common}.

\subsection{BLG EP angular factors}

For BLG, $F^{\rm BLG}=\cos^2\phi$ and
\begin{align}
 \int d\theta_kd\theta_p\,F^{\rm BLG}&=2\pi^2,\\
 \int d\theta_kd\theta_p\,q^2F^{\rm BLG}&=2\pi^2(k^2+p^2).
\end{align}
These averages lead to \(\mathcal{J}_0\) and \(\mathcal{J}_1\).  At $\hbar \omega=0$,
\begin{equation}
 \mathcal{J}_1(0;T_e)=8(k_{\rm B}T_{\rm e})^2
 \ln\left[2\cosh\left(\frac{\mu}{2k_{\rm B}T_{\rm e}}\right)\right],
\end{equation}
which reproduces the two-temperature BLG EP zero-frequency noise.

In the degenerate BLG EP limit, $|\mu|\gg k_{\rm B}T_{\rm e}$, the low-frequency ratio is
\begin{equation}
 \frac{S_{\rm EP}^{\rm BLG}(\omega)}
 {S_{\rm EP}^{\rm BLG}(0)}
 \simeq1+\frac{1}{12}\left(\frac{\hbar \omega}{k_{\rm B}T_{\rm e}}\right)^2
 +\frac{(\hbar \omega)^2}{16m v_{\rm s}^2|\mu|}+O((\hbar \omega)^4).
 \label{eq:app_blg_ep_quad}
\end{equation}

\section{Full on-shell MLG integral and numerical implementation}
\label{app:numerics}

\subsection{Dimensionless full intraband formula}

The full MLG benchmark keeps exact intraband kinematics for $\mu>0$.  Introduce $x=k/k_{\rm F}$, $y=q/k_{\rm F}$, $
 \eta=v_{\rm s}/v_{\rm F}$, $z=\hbar \omega/\mu$.
For frequency sign $\lambda=\pm1$ and process sign $s=\pm1$,
\begin{align}
 & u_{\lambda s}=\frac{\bar p_{\lambda s}^2-x^2-y^2}{2xy}, \quad \bar p_{\lambda s}=x+s\eta y-\lambda z,\\
 &F_{\lambda s}=\frac{1+\cos\phi_{\lambda s}}{2}, \quad \cos\phi_{\lambda s}=\frac{x+yu_{\lambda s}}{\bar p_{\lambda s}}.
\end{align}
The shell is allowed if $\bar p_{\lambda s}>0$ and $|u_{\lambda s}|<1$.  The phonon energy is $E_q=\eta\mu y$.  The phonon weight is $W_{+}=1+n_{\rm ph}$ for emission and $W_{-}=n_{\rm ph}$ for absorption. The exact spectrum implemented in the code is
\begin{widetext}
\begin{align}
 \frac{S(\omega;T_{\rm e},T_{\rm ph})}{A}
 &=\frac{g_dD^2k_{\rm F}^4}{32\pi^2\rho_{\rm MLG}v_{\rm s}\hbar v_{\rm F}}
\sum_{\lambda,s=\pm1}\int_0^{\infty}dx\int_0^{\infty}dy \frac{\bar p_{\lambda s}yF_{\lambda s}}
 {\sqrt{1-u_{\lambda s}^2}}
 f_{\rm e}(\mu\bar p_{\lambda s};T_{\rm e})[1-f_{\rm e}(\mu x;T_{\rm e})]\nonumber\\
 &\quad\times W_s(\eta\mu y;T_{\rm ph})
 [\lambda \hbar \omega-2s\eta\mu y]^2
 \Theta(\bar p_{\lambda s})\Theta(1-|u_{\lambda s}|).
 \label{eq:app_full_integral}
\end{align}
\end{widetext}
No BG small-$q$, EP classical-Bose, or quasi-elastic approximation enters Eq.~\eqref{eq:app_full_integral}.  The remaining restrictions are the linear MLG band, linear LA phonons, unscreened deformation potential, leading order in $D^2$, and the doped intraband channel.

\subsection{Numerical quadrature}

\begin{figure}[t]
 \centering
 \includegraphics[width=\columnwidth]{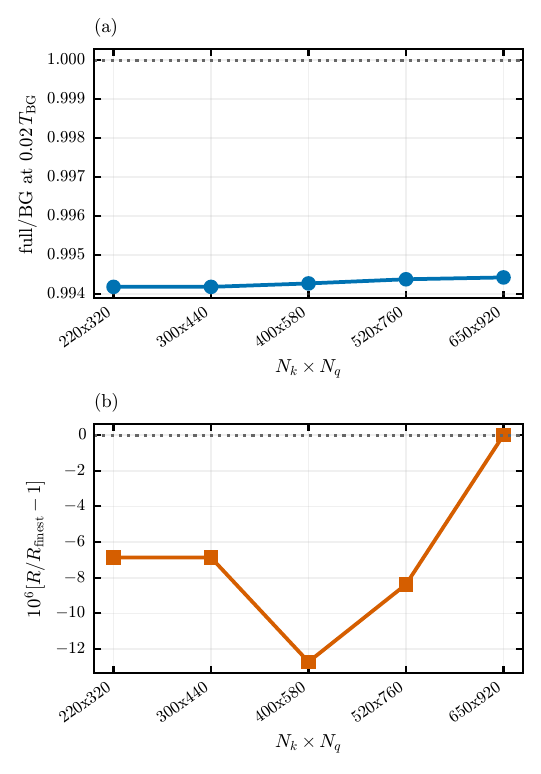}
\caption{Grid-convergence tests for the full MLG
on-shell integral, Eq.~\eqref{eq:app_full_integral}.
Here $N_k$ and $N_q$ are the numbers of integration
points for the electron and phonon wave-vector
magnitudes, respectively. Here $N_k$ and $N_q$ denote
the numbers of integration points for the electron
and phonon wave-vector magnitudes, respectively.
(a) Zero-frequency ratio
$S_{\rm full}(0)/S_{\rm BG}(0)$ at
$T_{\rm e}=T_{\rm ph}=0.02T_{\rm BG}$,
using Eq.~\eqref{eq:S0_MLG_BG} for the BG reference.
(b) Relative deviation
$10^6[R_{N_k,N_q}(\omega)/R_{\rm finest}(\omega)-1]$
at $T_{\rm e}=T_{\rm ph}=T_{\rm BG}$ and
$f=100\,\mathrm{GHz}$, expressed in parts per million.
The ratio $R$ is defined in
Eq.~\eqref{eq:R_definition}, and
$R_{\rm finest}$ uses the finest grid shown,
$(N_k,N_q)=(650,920)$.}
 \label{fig:convergence}
\end{figure}

The reduced BG, MLG EP, and BLG EP one-dimensional integrals are evaluated with Gauss--Legendre quadrature on finite intervals chosen so that the distribution functions suppress the neglected tails.  Stable series are used near zero argument for the Bose and electron phase-space functions. The full two-dimensional integral uses composite nonuniform grids.  The $k$ grid is refined around $k_{\rm F}$ over the thermal width
$\delta k\sim k_{\rm B}T_{\rm e}/\hbar v_{\rm F}$
while the $q$ grid resolves both the thermal phonon scale $q_{\rm T}=k_{\rm B}T_{\rm ph}/(\hbar v_{\rm s})$ and the large-$q$ EP region.  Kinematically forbidden points are removed by the two step functions in Eq.~\eqref{eq:app_full_integral}.  The square-root edge is regulated only at machine-level distance from the boundary and Convergence is checked by increasing $N_k$ and $N_q$,
the numbers of integration points in the electron
and phonon wave-vector magnitudes, respectively.
Fig.~\ref{fig:convergence} shows two representative
tests. Panel (a) compares the converged full
zero-frequency noise at $T=0.02T_{\rm BG}$ with
the analytic BG expression.
Panel (b) tests the stability of
$R(\omega)=S(\omega)/S(0)$ at $T=T_{\rm BG}$
and $f=100\,\mathrm{GHz}$ by plotting its relative
deviation from the finest-grid result in parts
per million.
The validation script also verifies the
zero-frequency normalizations and the analytic
quadratic coefficients.

\clearpage
\bibliography{references}

@article{Hwang2008,
  author = {Hwang, E. H. and Das Sarma, S.},
  title = {Acoustic phonon scattering limited carrier mobility in two-dimensional extrinsic graphene},
  journal = {Phys. Rev. B},
  volume = {77},
  pages = {115449},
  year = {2008},
  doi = {10.1103/PhysRevB.77.115449}
}

@article{Kubakaddi2009,
  author = {Kubakaddi, S. S.},
  title = {Interaction of massless Dirac electrons with acoustic phonons in graphene at low temperatures},
  journal = {Phys. Rev. B},
  volume = {79},
  pages = {075417},
  year = {2009},
  doi = {10.1103/PhysRevB.79.075417}
}

@article{Bistritzer2009,
  author = {Bistritzer, R. and MacDonald, A. H.},
  title = {Electronic Cooling in Graphene},
  journal = {Phys. Rev. Lett.},
  volume = {102},
  pages = {206410},
  year = {2009},
  doi = {10.1103/PhysRevLett.102.206410}
}

@article{Tse2009,
  author = {Tse, Wang-Kong and Das Sarma, S.},
  title = {Energy relaxation of hot Dirac fermions in graphene},
  journal = {Phys. Rev. B},
  volume = {79},
  pages = {235406},
  year = {2009},
  doi = {10.1103/PhysRevB.79.235406}
}

@article{Viljas2010,
  author = {Viljas, J. K. and Heikkil\"a, T. T.},
  title = {Electron-phonon heat transfer in monolayer and bilayer graphene},
  journal = {Phys. Rev. B},
  volume = {81},
  pages = {245404},
  year = {2010},
  doi = {10.1103/PhysRevB.81.245404}
}

@article{Song2012,
  author = {Song, Justin C. W. and Reizer, Michael Y. and Levitov, Leonid S.},
  title = {Disorder-Assisted Electron-Phonon Scattering and Cooling Pathways in Graphene},
  journal = {Phys. Rev. Lett.},
  volume = {109},
  pages = {106602},
  year = {2012},
  doi = {10.1103/PhysRevLett.109.106602}
}

@article{Fong2012,
  author = {Fong, Kin Chung and Schwab, K. C.},
  title = {Ultrasensitive and Wide-Bandwidth Thermal Measurements of Graphene at Low Temperatures},
  journal = {Phys. Rev. X},
  volume = {2},
  pages = {031006},
  year = {2012},
  doi = {10.1103/PhysRevX.2.031006}
}

@article{Betz2012,
  author = {Betz, A. C. and Vialla, F. and Brunel, D. and Voisin, C. and Picher, M. and Cavanna, A. and Madouri, A. and F\`eve, G. and Berroir, J.-M. and Pla\c{c}ais, B. and Pallecchi, E.},
  title = {Hot Electron Cooling by Acoustic Phonons in Graphene},
  journal = {Phys. Rev. Lett.},
  volume = {109},
  pages = {056805},
  year = {2012},
  doi = {10.1103/PhysRevLett.109.056805}
}

@article{Betz2013,
  author = {Betz, A. C. and Jhang, S. H. and Pallecchi, E. and Ferreira, R. and F\`eve, G. and Berroir, J.-M. and Pla\c{c}ais, B.},
  title = {Supercollision cooling in undoped graphene},
  journal = {Nat. Phys.},
  volume = {9},
  pages = {109--112},
  year = {2013},
  doi = {10.1038/nphys2494}
}

@article{Graham2013,
  author = {Graham, M. W. and Shi, S.-F. and Ralph, D. C. and Park, J. and McEuen, P. L.},
  title = {Photocurrent measurements of supercollision cooling in graphene},
  journal = {Nat. Phys.},
  volume = {9},
  pages = {103--108},
  year = {2013},
  doi = {10.1038/nphys2493}
}

@article{Callen1951,
  author = {Callen, H. B. and Welton, T. A.},
  title = {Irreversibility and Generalized Noise},
  journal = {Phys. Rev.},
  volume = {83},
  pages = {34--40},
  year = {1951},
  doi = {10.1103/PhysRev.83.34}
}

@article{Kubo1957,
  author = {Kubo, R.},
  title = {Statistical-Mechanical Theory of Irreversible Processes. I. General Theory and Simple Applications to Magnetic and Conduction Problems},
  journal = {J. Phys. Soc. Jpn.},
  volume = {12},
  pages = {570--586},
  year = {1957},
  doi = {10.1143/JPSJ.12.570}
}

@article{Averin2010,
  author = {Averin, D. V. and Pekola, J. P.},
  title = {Violation of the Fluctuation-Dissipation Theorem in Time-Dependent Mesoscopic Heat Transport},
  journal = {Phys. Rev. Lett.},
  volume = {104},
  pages = {220601},
  year = {2010},
  doi = {10.1103/PhysRevLett.104.220601}
}

@article{Pekola2018,
  author = {Pekola, Jukka P. and Karimi, Bayan},
  title = {Quantum Noise of Electron-Phonon Heat Current},
  journal = {J. Low Temp. Phys.},
  volume = {191},
  pages = {373--379},
  year = {2018},
  doi = {10.1007/s10909-018-1854-y}
}

@article{Giazotto2006,
  author = {Giazotto, F. and Heikkil\"a, T. T. and Luukanen, A. and Savin, A. M. and Pekola, J. P.},
  title = {Opportunities for mesoscopics in thermometry and refrigeration: Physics and applications},
  journal = {Rev. Mod. Phys.},
  volume = {78},
  pages = {217--274},
  year = {2006},
  doi = {10.1103/RevModPhys.78.217}
}

@article{Gasparinetti2015,
  author = {Gasparinetti, S. and Viisanen, K. L. and Saira, O.-P. and Faivre, T. and Arzeo, M. and Meschke, M. and Pekola, J. P.},
  title = {Fast Electron Thermometry for Ultrasensitive Calorimetric Detection},
  journal = {Phys. Rev. Applied},
  volume = {3},
  pages = {014007},
  year = {2015},
  doi = {10.1103/PhysRevApplied.3.014007}
}

@article{Saira2016,
  author = {Saira, O.-P. and Zgirski, M. and Viisanen, K. L. and Golubev, D. S. and Pekola, J. P.},
  title = {Dispersive Thermometry with a Josephson Junction Coupled to a Resonator},
  journal = {Phys. Rev. Applied},
  volume = {6},
  pages = {024005},
  year = {2016},
  doi = {10.1103/PhysRevApplied.6.024005}
}

@article{Novoselov2004,
  author = {Novoselov, K. S. and Geim, A. K. and Morozov, S. V. and Jiang, D. and Zhang, Y. and Dubonos, S. V. and Grigorieva, I. V. and Firsov, A. A.},
  title = {Electric field effect in atomically thin carbon films},
  journal = {Science},
  volume = {306},
  number = {5696},
  pages = {666--669},
  year = {2004},
  doi = {10.1126/science.1102896}
}

@article{Novoselov2005,
  author = {Novoselov, K. S. and Geim, A. K. and Morozov, S. V. and Jiang, D. and Katsnelson, M. I. and Grigorieva, I. V. and Dubonos, S. V. and Firsov, A. A.},
  title = {Two-dimensional gas of massless Dirac fermions in graphene},
  journal = {Nature},
  volume = {438},
  pages = {197--200},
  year = {2005},
  doi = {10.1038/nature04233}
}

@article{McCann2006,
  author = {McCann, Edward and Fal'ko, Vladimir I.},
  title = {Landau-Level Degeneracy and Quantum Hall Effect in a Graphite Bilayer},
  journal = {Phys. Rev. Lett.},
  volume = {96},
  pages = {086805},
  year = {2006},
  doi = {10.1103/PhysRevLett.96.086805}
}

@article{McCann2013,
  author = {McCann, Edward and Koshino, Mikito},
  title = {The electronic properties of bilayer graphene},
  journal = {Rep. Prog. Phys.},
  volume = {76},
  pages = {056503},
  year = {2013},
  doi = {10.1088/0034-4885/76/5/056503}
}

@article{Efetov2010,
  author = {Efetov, Dmitri K. and Kim, Philip},
  title = {Controlling Electron-Phonon Interactions in Graphene at Ultrahigh Carrier Densities},
  journal = {Phys. Rev. Lett.},
  volume = {105},
  pages = {256805},
  year = {2010},
  doi = {10.1103/PhysRevLett.105.256805}
}

@article{Borysenko2010,
  author = {Borysenko, K. M. and Mullen, J. T. and Barry, E. A. and Paul, S. and Semenov, Y. G. and Zavada, J. M. and Buongiorno Nardelli, M. and Kim, K. W.},
  title = {First-principles analysis of electron-phonon interactions in graphene},
  journal = {Phys. Rev. B},
  volume = {81},
  pages = {121412(R)},
  year = {2010},
  doi = {10.1103/PhysRevB.81.121412}
}

@article{Borysenko2011,
  author = {Borysenko, K. M. and Mullen, J. T. and Li, X. and Semenov, Y. G. and Zavada, J. M. and Buongiorno Nardelli, M. and Kim, K. W.},
  title = {Electron-phonon interactions in bilayer graphene},
  journal = {Phys. Rev. B},
  volume = {83},
  pages = {161402(R)},
  year = {2011},
  doi = {10.1103/PhysRevB.83.161402}
}

@article{Min2011,
  author = {Min, Hongki and Hwang, E. H. and Das Sarma, S.},
  title = {Chirality-dependent phonon-limited resistivity in multiple layers of graphene},
  journal = {Phys. Rev. B},
  volume = {83},
  pages = {161404(R)},
  year = {2011},
  doi = {10.1103/PhysRevB.83.161404}
}

@article{Kaasbjerg2012,
  author = {Kaasbjerg, Kristen and Thygesen, Kristian S. and Jacobsen, Karsten W.},
  title = {Unraveling the acoustic electron-phonon interaction in graphene},
  journal = {Phys. Rev. B},
  volume = {85},
  pages = {165440},
  year = {2012},
  doi = {10.1103/PhysRevB.85.165440}
}

@article{Sohier2014,
  author = {Sohier, Thibault and Calandra, Matteo and Park, Cheol-Hwan and Bonini, Nicola and Marzari, Nicola and Mauri, Francesco},
  title = {Phonon-limited resistivity of graphene by first-principles calculations: Electron-phonon interactions, strain-induced gauge field, and Boltzmann equation},
  journal = {Phys. Rev. B},
  volume = {90},
  pages = {125414},
  year = {2014},
  doi = {10.1103/PhysRevB.90.125414}
}

@article{ChenClerk2012,
  author = {Chen, Wei and Clerk, Aashish A.},
  title = {Electron-phonon mediated heat flow in disordered graphene},
  journal = {Phys. Rev. B},
  volume = {86},
  pages = {125443},
  year = {2012},
  doi = {10.1103/PhysRevB.86.125443}
}

@article{Baker2012,
  author = {Baker, A. M. R. and Alexander-Webber, J. A. and Altebaeumer, T. and Nicholas, R. J.},
  title = {Energy relaxation for hot Dirac fermions in graphene and breakdown of the quantum Hall effect},
  journal = {Phys. Rev. B},
  volume = {85},
  pages = {115403},
  year = {2012},
  doi = {10.1103/PhysRevB.85.115403}
}

@article{Baker2013,
  author = {Baker, A. M. R. and Alexander-Webber, J. A. and Altebaeumer, T. and McMullan, S. D. and Janssen, T. J. B. M. and Tzalenchuk, A. and Lara-Avila, S. and Kubatkin, S. and Yakimova, R. and Lin, C.-T. and Li, L.-J. and Nicholas, R. J.},
  title = {Energy loss rates of hot Dirac fermions in epitaxial, exfoliated, and chemical-vapor-deposited graphene},
  journal = {Phys. Rev. B},
  volume = {87},
  pages = {045414},
  year = {2013},
  doi = {10.1103/PhysRevB.87.045414}
}

@article{McKitterick2016,
  author = {McKitterick, Christopher B. and Prober, Daniel E. and Rooks, Michael J.},
  title = {Electron-phonon cooling in large monolayer graphene devices},
  journal = {Phys. Rev. B},
  volume = {93},
  pages = {075410},
  year = {2016},
  doi = {10.1103/PhysRevB.93.075410}
}

@article{Laitinen2014,
  author = {Laitinen, Antti and Oksanen, Mika and Fay, Aur\'elien and Cox, Daniel and Tomi, Matti and Virtanen, Pauli and Hakonen, Pertti J.},
  title = {Electron-Phonon Coupling in Suspended Graphene: Supercollisions by Ripples},
  journal = {Nano Lett.},
  volume = {14},
  pages = {3009--3013},
  year = {2014},
  doi = {10.1021/nl404258a}
}

@article{Borzenets2013,
  author = {Borzenets, I. V. and Coskun, U. C. and Mebrahtu, H. T. and Bomze, Yu. V. and Smirnov, A. I. and Finkelstein, G.},
  title = {Phonon Bottleneck in Graphene-Based Josephson Junctions at Millikelvin Temperatures},
  journal = {Phys. Rev. Lett.},
  volume = {111},
  pages = {027001},
  year = {2013},
  doi = {10.1103/PhysRevLett.111.027001}
}

@article{Yan2012,
  author = {Yan, Jun and Kim, M.-H. and Elle, J. A. and Sushkov, A. B. and Jenkins, G. S. and Milchberg, H. M. and Fuhrer, M. S. and Drew, H. D.},
  title = {Dual-gated bilayer graphene hot-electron bolometer},
  journal = {Nat. Nanotechnol.},
  volume = {7},
  pages = {472--478},
  year = {2012},
  doi = {10.1038/nnano.2012.88}
}

@article{Vora2012,
  author = {Vora, Heli and Kumaravadivel, Piranavan and Nielsen, Bent and Du, Xu},
  title = {Bolometric response in graphene based superconducting tunnel junctions},
  journal = {Appl. Phys. Lett.},
  volume = {100},
  pages = {153507},
  year = {2012},
  doi = {10.1063/1.3703117}
}

@article{Fong2013,
  author = {Fong, Kin Chung and Wollman, Emma E. and Ravi, Harish and Chen, Wei and Clerk, Aashish A. and Shaw, M. D. and Leduc, H. G. and Schwab, K. C.},
  title = {Measurement of the Electronic Thermal Conductance Channels and Heat Capacity of Graphene at Low Temperature},
  journal = {Phys. Rev. X},
  volume = {3},
  pages = {041008},
  year = {2013},
  doi = {10.1103/PhysRevX.3.041008}
}

@article{Kokkoniemi2020,
  author = {Kokkoniemi, Roope and Girard, Jean-Philippe and Hazra, Dibyendu and Laitinen, Antti and Govenius, Joonas and Lake, Russell E. and Sallinen, Iiro and Vesterinen, Visa and Partanen, Matti and Tan, J. Y. and Chan, K. W. and Tan, K. Y. and Hakonen, Pertti J. and M\"ott\"onen, Mikko},
  title = {Bolometer operating at the threshold for circuit quantum electrodynamics},
  journal = {Nature},
  volume = {586},
  pages = {47--51},
  year = {2020},
  doi = {10.1038/s41586-020-2753-3}
}

@article{Karimi2021,
  author = {Karimi, Bayan and Pekola, Jukka P.},
  title = {Downconversion of quantum fluctuations of photonic heat current in a circuit},
  journal = {Phys. Rev. B},
  volume = {104},
  pages = {165418},
  year = {2021},
  doi = {10.1103/PhysRevB.104.165418}
}

@article{KaasbjergCooling2014,
  author = {Kaasbjerg, Kristen and Bhargavi, K. S. and Kubakaddi, S. S.},
  title = {Hot-electron cooling by acoustic and optical phonons in monolayers of {MoS2} and other transition-metal dichalcogenides},
  journal = {Phys. Rev. B}, volume = {90}, pages = {165436}, year = {2014},
  doi = {10.1103/PhysRevB.90.165436}
}

@article{IdrisovThermalDrag2022,
  author = {Idrisov, Edvin G. and Levkivskyi, Ivan P. and Sukhorukov, Eugene V.},
  title = {Thermal drag effect in quantum Hall circuits},
  journal = {Phys. Rev. B}, volume = {106}, pages = {L121405}, year = {2022},
  doi = {10.1103/PhysRevB.106.L121405}
}

@article{IdrisovAnyonicHeat2024,
  author = {Idrisov, Edvin G. and Ekstr{\"o}m, Johan},
  title = {Quantization of the anyonic spectral density of heat currents},
  journal = {Phys. Rev. B}, volume = {110}, pages = {L041402}, year = {2024},
  doi = {10.1103/PhysRevB.110.L041402}
}

\end{document}